\documentclass[%
reprint,
onecolumn,
superscriptaddress,
amsmath,amssymb,
aps,
pra,
showkeys
]{revtex4-2}

\usepackage{multirow}
\usepackage{graphicx}
\usepackage{dcolumn}
\usepackage{bm}
\usepackage[colorlinks=true,
            linkcolor=red,
            citecolor=blue,
            urlcolor=blue]{hyperref}
\usepackage[mathlines]{lineno}
\usepackage{mathtools}
\usepackage[dvipsnames,table]{xcolor}
\usepackage{longtable}
\usepackage{makecell}

\newcommand{\TT}[1]{{#1}}
\newcommand{\TTT}[1]{{#1}}

\begin{document}

\title{Quantum correlations and spatial localization in trapped one-dimensional ultra-cold Bose-Bose-Bose mixtures}
\author{Tran Duong Anh-Tai\footnote{Author to whom any correspondence should be addressed}} 
\email{tai.tran@oist.jp,tai.d.tran-1@ou.edu}
\affiliation{Quantum Systems Unit, OIST Graduate University, Onna, Okinawa 904-0495, Japan}
\affiliation{Homer L. Dodge Department of Physics and Astronomy, The University of Oklahoma, 440 W. Brooks Street, Norman, Oklahoma 73019, USA}
\affiliation{Center for Quantum Research and Technology, The University of Oklahoma, 440 W. Brooks Street, Norman, Oklahoma 73019, USA}
\author{Miguel A. García-March} 
\email{garciamarch@mat.upv.es}
\affiliation{Instituto Universitario de Matem\'atica Pura y Aplicada, Universitat Polit\`ecnica de Val\`encia, E-46022 Val\`encia, Spain}
\author{Thomas Busch}
\affiliation{Quantum Systems Unit, OIST Graduate University, Onna, Okinawa 904-0495, Japan} 
\author{Thomás Fogarty} 
\email{thomas.fogarty@oist.jp}
\affiliation{Quantum Systems Unit, OIST Graduate University, Onna, Okinawa 904-0495, Japan}
\date{\today}

\begin{abstract}
We systematically investigate and illustrate the complete ground-state phase diagram for a one-dimensional, three-species mixture of a few repulsively interacting bosons trapped harmonically. To numerically obtain the solutions to the many-body Schr\"{o}dinger equation, we employ the improved Exact Diagonalization method [T. D. Anh-Tai {\it et al.}, SciPost Physics 15, 048 (2023)], which is capable of treating strongly-correlated few-body systems from first principles in an efficiently truncated Hilbert space. We present our comprehensive results for all possible combinations of intra- and interspecies interactions in the extreme limits that are either the ideal limit ($g=0$) or close to the hard-core limit ($g\to\infty$). These results show the emergence of unique ground-state properties related to correlations, coherence and spatial localization stemming from strongly repulsive interactions.
\end{abstract}

\maketitle

\section{Introduction}
\label{sec:intro}
Over the past three decades, ultra-cold atomic gases have been an excellent and unique platform to explore the fascinating physics of complex many-body quantum systems in a very clean setting with high degrees of control. Moreover, they possess a great number of promising applications for quantum technologies such as quantum simulators \cite{bloch2012quantum,karman2024ultracold,cornish2024quantum}, and quantum metrology \cite{demille2024quantum}. Although the physics of weakly-interacting ultra-cold bosonic gases is important and well understood within the framework of Gross-Pitaevskii mean-field theory \cite{pethick2008Bose,pitaevskii2016bose}, investigating systems of few bosons, fermions, and mixtures thereof in low dimensional space, where correlations are of importance, is an equally important and intriguing area of research \cite{sowinski2019one,mistakidis2023few}. Recent advances in laser cooling techniques and quantum optics have made it feasible to create strongly-correlated systems in low dimensions with well-defined particle numbers in \TT{several laboratories, with single \cite{kinoshita2004observation,kinoshita2005local,paredes2004tonks,serwane2011deterministic,wenz2013few,murmann2015two} and multi-component systems being realized \cite{zurn2012fermionization,zurn2013pairing,pagano2014one,murmann2015antiferromagnetic}, and even being able to measure entanglement in few-body systems \cite{bergschneider2019experimental,BECKER2020}. These experiments and the degree of control available in modern setups have} stimulated extensive beyond-mean-field studies of few-body one- and two-component systems in parallel to mean-field studies.

When the repulsive interaction strength in a one-dimensional single-species bosonic gas is varied from being weak to being strong, the system undergoes a transition from condensation to fermionization \cite{zollner2006correlations,deuretzbacher2007evolution}. In the infinitely repulsive limit, the bosonic system can be mapped to a non-interacting, spin-polarized Fermi gas and its wavefunction can be analytically obtained by the Bose-Fermi mapping theorem \cite{girardeau1960relationship}. This is referred to as the Tonks-Girardeau hard-core limit, which has been experimentally realized \cite{kinoshita2004observation,paredes2004tonks}. Meanwhile, binary mixtures exhibit additional phenomena due to the presence of the interspecies interaction or different particle statistics. For instance, when the interspecies coupling strengths are large, binary bosonic mixtures can exhibit a phase-separated state \cite{myatt1997production,cazalilla2003instabilities,alon2006demixing,mishra2007phase,garcia2013quantum} or form a composite-fermionzation phase \cite{hao2009ground1,sascha2008composite,garcia2013sharp} depending on the intraspecies interactions being strong or weak, respectively. Similar phases can appear in binary mixtures of few particles, with the ground-state properties having been fully explored in Ref.~\cite{garcia2014quantum}. Furthermore, when considering weaker interactions away from the integrable hard-core and ideal BEC limits the system is rather complex and can display strong signatures of quantum chaos due to the abundance of avoided crossings in the energy spectrum \cite{chen2021impurity,anhtai2023quantum}. Importantly, binary few-body mixtures have also emerged as an excellent platform for gaining deeper insights into impurity physics \cite{garcia2016entanglement,2018DehkharghaniPRL,2022TheelPRA,2020MistakidisPRR,mistakidis2019repulsive,Tai2024inprepPRX} and few-body quantum droplets \cite{chergui2023superfluid,mistakidis2021formation,englezos2023particle,englezos2023correlated}, as correlation effects can be more easily assessed due to access to the full quantum many-body state.  

Extending studies to triple-species mixtures, which possess an even larger parameter space compared to single and binary mixtures, is therefore likely to unveil even richer physics and function as a guide to future experiments. Although it is computationally challenging to accurately solve the many-body Schr\"{o}dinger equation in continuous 1D space due to the large Hilbert space, efficient numerical tools have recently been developed for this purpose such as the multi-layer multi-configuration time-dependent Hartree method for mixtures of identical particles~\cite{cao2017unified,cao2013multi,kronke2013non} or the improved Exact Diagonalization method~\cite{anhtai2023quantum}. So far, the studies on three-species few-particle systems have mainly focused on correlations and entanglement between two distinguishable impurities coupled to a quantum few-boson bath~\cite{2021KeilerPRA,theel2024crossover} and on engineering strongly-correlated atomic Bell states~\cite{Tai2024inprepPRX}. For a more systematic approach we will in the following explore the ground-state properties, including the correlations, coherence and self-organization, across all possible interaction regimes for a three-species mixture of a few bosons confined in a one-dimensional harmonic trap from first principles.

The main goal of this work is to explore and illustrate the complete ground-state phase diagram of the system when the intra- and interspecies interactions are either in the ideal limit ($g=0$) or close to the hard-core limit ($g\to\infty$). For this we use the one-body density distribution function to characterize the spatial localization, the reduced one- and two-body density matrices, and the bipartite mutual information as the indicators for quantum correlations and coherence. As one of the main results, we find that the ground-state phases of the system with respect to all possible combinations of the interaction strengths can be classified into two groups. The first group consists of phases that exhibit correlations in only one or two species and this group is well-studied theoretically in previous works. In the second group all three species are coupled and hence exhibit interesting results which are unique to three-species bosonic mixtures. We therefore focus on the second group and concisely explore the static ground-state properties of all possible combinations of interaction strengths in this group. Additionally, we systematically investigate the correlations and coherence properties for a representative example where two intraspecies coupling strengths vary between the ideal BEC limit and the hard-core limit, thereby connecting two of the limiting phases and \TT{demonstrating} that interesting effects can \TT{also be} observed in the moderate interaction regime. Our numerical approach to the solution of the many-body Schr\"{o}dinger equation is based on the improved Exact Diagonalization method~\cite{anhtai2023quantum}, which grants us access to the correlated ground-state wavefunction of the system, and the above-mentioned quantities of interest with a reasonable computational cost. \TT{Our results therefore provide insights into correlation effects in complex many-body quantum systems in the case of strong particle-particle interactions, that are relevant to future experiments in strongly-correlated multi-species ultra-cold quantum gases.}

The manuscript is organized as follows: Section~\ref{sec:model} introduces the Hamiltonian of our model, the ab initio method employed for the numerical solutions of the many-body Schr\"{o}dinger equation, and the quantities of interest characterizing quantum correlations, coherence and spatial self-organization. In Section~\ref{sec:discussion} we present the main findings related to the static ground-state properties of three unique tri-correlated phases and one representative connection between extreme states. Section~\ref{sec:conclusion} presents the conclusions and outlook. Finally, for completeness, we discuss the remaining tri-correlated cases, which can be seen as extensions of smaller systems, in Appendix~\ref{appendix}.
 
\section{Model, Methodology, and Quantities of interest}
\label{sec:model}
\subsection{Model}
We consider a three-species mixture of repulsively interacting bosonic atoms trapped in a one-dimensional parabolic potential with frequency $\omega$. We assume that each component $\sigma\in\{\text{A,B,C}\}$ has a minimal, but well-defined particle number, $N_\sigma = 2$, and that all masses are equal, $m_\sigma = m$. Since we only consider systems at low temperatures, the two-body scattering potential is well captured by a $s$-wave pseudo-potential that is usually modeled by a $\delta$-function~\cite{olshanii1998atomic}. Hereafter, we use harmonic oscillator units to rescale the many-body Hamiltonian which means that all lengths, energies, and coupling strengths will be given in terms of $\sqrt{\hbar/(m\omega)}$, $\hbar \omega$, and $\sqrt{\hbar^3\omega/m}$, respectively. The dimensionless Hamiltonian describing our system reads
\begin{align}
	\label{eq:full_hamiltonian_first}
	\hat{H} = \sum_{\sigma} \hat{H}^{\sigma} + \sum_{\sigma\neq \delta} \hat{W}^{\sigma\delta},
\end{align}
where $\hat{H}^{\sigma}$ denotes the $\sigma$-species Hamiltonian, while $\hat{W}^{\sigma\delta}$ describes the interactions between two species $\sigma$ and $\delta$. Explicitly they are given as
\begin{align}
    \hat{H}^{\sigma} &= \sum_{i=1}^{N_\sigma} \left[-\dfrac{1}{2}\dfrac{d^2}{d(x^\sigma_i)^2} + \dfrac{1}{2}(x^\sigma_i)^2 + g_\sigma\sum_{i<j}\delta(x^\sigma_i-x^\sigma_j)\right], \\ 
    \hat{W}^{\sigma\delta} &= g_{\sigma\delta} \sum_{i=1}^{N_\sigma}\sum_{j=1}^{N_\delta}\delta(x^\sigma_i-x^{\delta}_j).
\end{align}
Here the terms $g_\sigma$ and $g_{\sigma\delta}$ correspond to the intra- and interspecies coupling strengths respectively, which can be experimentally tuned from the ideal BEC limit $(g = 0)$ to the hard-core Tonks-Girardeau (TG) limit $(g \to \infty)$ by using Feshbach \cite{chin2010feshbach} or confinement-induced resonances \cite{haller2010confinement}. \TT{Since numerical calculations cannot directly handle $g = \infty$, we approximate the hard-core TG limit with $g = 20$ which has been shown in previous works to give results that are sufficiently close to infinite interactions \cite{hao2009ground1,garcia2013quantum,garcia2013sharp,garcia2014quantum,anhtai2023quantum}.}

Given that we consider a three-species mixture with elastic two-body collisions only, the system is described by six coupling strengths that explicitly are $g_{\rm{A}}$, $g_{\rm{B}}$, $g_{\rm{C}}$, $g_{\rm{AB}}$, $g_{\rm{AC}}$, and $g_{\rm{BC}}$, and since we are only interested in the limits when these strengths are either $g=0$ or $g\to \infty$, all possible combinations of the six coupling strengths result in a total of 64 cases to be considered. However, due to the assumption of equal masses and particle numbers in each species these reduce to \TT{20} distinct ones. \TT{We remark that some of the distinct states can be straightforwardly understood from the known solutions of the single and binary mixtures, whenever one or two components are decoupled. In particular, in the absence of all interspecies interactions, $g_{\rm{AB}}=g_{\rm{AC}}=g_{\rm{BC}}=0$, the system  lacks any interspecies correlations leading to the fact that its ground state can be simply factorized as $\Phi_\text{A}\otimes\Phi_\text{B}\otimes\Phi_\text{C}$ with $\Phi_{\sigma}$ being either in the BEC or TG limit. When only one of the interspecies interactions is present, i.e. $g_\text{AB}\rightarrow\infty$, the system still partly factorizes as $\Phi_\text{AB}\otimes\Phi_\text{C}$, with the uncoupled species $\text{C}$ ($\Phi_\text{C}$) and the bi-correlated state of $\text{A}$ and $\text{B}$ ($\Phi_\text{AB}$). For such cases the bi-correlated phases have been described in detail in Ref.~\cite{garcia2014quantum}, where the appearance of phase separation~\cite{garcia2013quantum,garcia2014quantum}, composite fermionization~\cite{hao2009ground1,sascha2008composite,garcia2013sharp} and full fermionization~\cite{girardeau2007soluble} was confirmed.}

\begin{figure}
    \centering
    \includegraphics[width=0.95\linewidth]{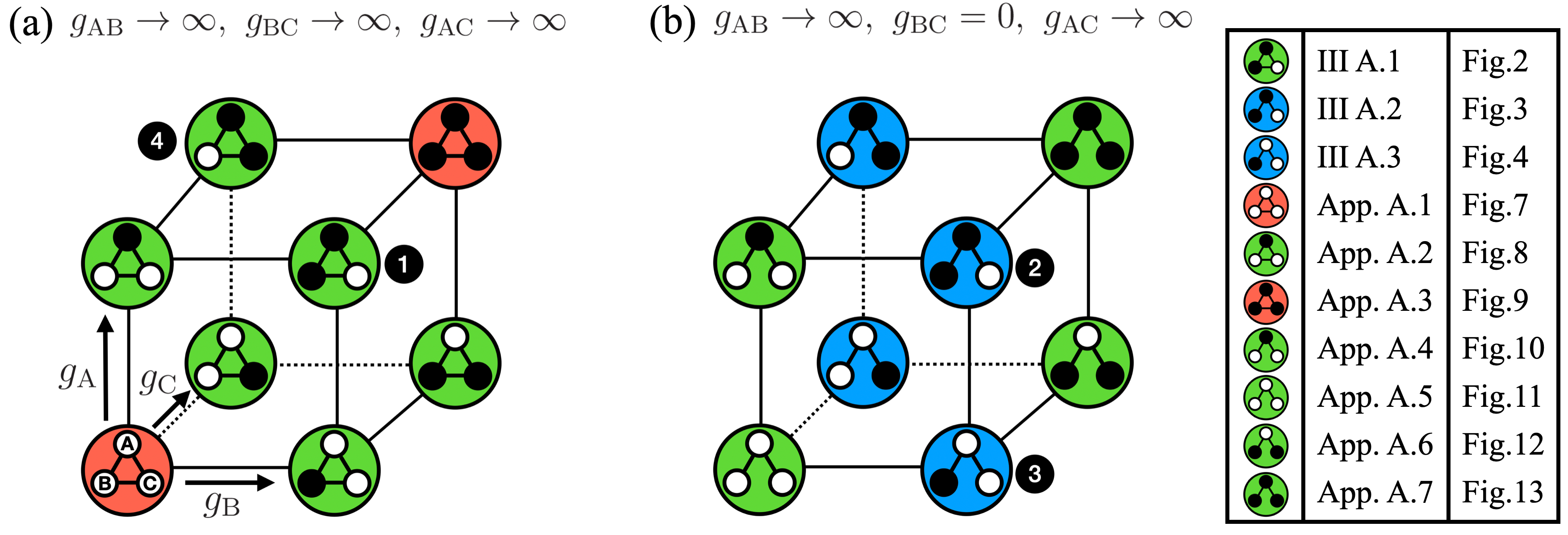}
    \caption{\TT{Tri-correlated states represented in two cubes spanning the intraspecies interactions, $g_\text{A}$, $g_\text{B}$ and $g_\text{C}$. (a) Phase cube for isotropic interspecies interactions $g_\text{AB}\rightarrow\infty$, $g_\text{BC}\rightarrow\infty$ and $g_\text{BC}\rightarrow\infty$, and (b) phase cube for anisotropic interspecies interactions $g_\text{AB}\rightarrow\infty$, $g_\text{BC}=0$ and $g_\text{BC}\rightarrow\infty$.
    The vertices of each cube visualize the state of the tri-component system, with the small circles denoting the intraspecies interactions (white for non-interacting and black for infinite repulsive interactions), while the lines connecting circles indicate the presence infinite repulsive interspecies interactions between the two components. The red phases highlight states which are invariant under exchange of any of the three components, green phases are invariant following exchange of two species and blue phases that are not invariant under any species exchange. The numbered vertices are discussed in detail in the main text, and \TTT{the legend indicates the corresponding section and figure where the phases are discussed.} } }
    \label{fig:ice-cube}
\end{figure}

\TT{In the following we focus on the intriguing tri-correlated states, in which each species is interacting with one or more of the other components. For this one can visualize the appearance of the different phases of the total system by two cubes, shown in Fig.~\ref{fig:ice-cube}. For the cube shown in panel (a) all species interact with each other ($g_\text{AB}\rightarrow\infty$, $g_\text{BC}\rightarrow\infty$ and $g_\text{AC}\rightarrow\infty$) and for the cube shown in panel (b) only two of the interspecies interactions are finite ($g_\text{AB}\rightarrow\infty$, $g_\text{BC}=0$ and $g_\text{AC}\rightarrow\infty$). The dimensions in each cube span the intraspecies interactions, $g_\text{A}$, $g_\text{B}$ and $g_\text{C}$ and the vertices represent the different phases. For each phase we visualize the intraspecies interactions in the individual components, A, B and C, with three small circles with their color being white indicating no intraspecies interactions ($g_\sigma=0$) and black indicating infinite repulsive intraspecies interactions ($g_\sigma\rightarrow\infty$). If two species interact with an infinitely repulsive interaction, we connect the corresponding two small circles with a line, whereas this line is absent if the interspecies interaction is zero.  Finally, the different colors of the vertices themselves indicate how many species are invariant after exchange with another species. Red indicates that the system is invariant under exchange of any of the three species, as all intraspecies interactions and all interspecies interactions are the same. Green indicates that only two species are invariant, while blue indicates that the system is not invariant under any exchange of components.} 

When considering the cube with isotropic interspecies interactions in panel (a) we find four unique phases, two of which are three species invariant \TTT{(red color)}. These are the three-species analogs of known phases in two-component systems \cite{garcia2014quantum}: composite fermionization (infinite interspecies coupling but zero intrapsecies coupling) and full fermionization (all interactions are infinite), which are discussed in the appendices Sec.~\ref{TCF} and Sec.~\ref{TFF} respectively. The two other unique phases in this cube are two species invariant \TTT{(green color) and each contains three different permutations that are related to one another through reflection symmetry. The phase labeled \textbf{1} in Fig.~\ref{fig:ice-cube}(a) is discussed in detail in Sec.~\ref{FPS} while the other phase is discussed in Appendix~\ref{DICFPS}.}

The cube with anisotropic interspecies interactions shown in panel (b) has one interspecies interaction being zero and therefore results in nontrivial phases which have no two component analog. We note that permutations of the interspecies interactions result in the same phases but with the vertices of the cube swapped. In this cube we find six unique phases, four of which are invariant under two species exchange \TTT{(green color)}, specifically whenever species B and C have the same intraspecies interactions $g_\text{B}=g_\text{C}$, which we detail in appendices~\ref{PS}, \ref{CIBI}, \ref{CIBIII} and \ref{CIABII}. Otherwise, when $g_\text{B}\neq g_\text{C}$ we have two phases which are not invariant under any species exchange \TTT{(blue color)}, \TTT{each of which have two permutations which are related via a reflection symmetry. The states which are labeled \textbf{2} and \textbf{3} in Fig.~\ref{fig:ice-cube}(b) are discussed in Sec.~\ref{CIABI} and Sec.~\ref{CIBII}.}

\subsection{Methodology}
\label{app:method}
While accurately solving the many-body Schr\"{o}dinger equation is paramount when exploring correlations in complex interacting quantum systems, it is a computational challenge due to the usually large Hilbert space. In this work we employ the improved Exact Diagonalization method \cite{anhtai2023quantum} to numerically diagonalize the Hamiltonian \eqref{eq:full_hamiltonian_first} and obtain the ground state of the system in the different parameter regimes. Since we are treating quantum systems consisting of identical particles, it is convenient to rewrite the Hamiltonian \eqref{eq:full_hamiltonian_first} in the second-quantized formalism by introducing the $\sigma$-component bosonic annihilation operator $\hat{a}_{\sigma,k}$ as
\begin{equation}
    \hat{a}_{\sigma,k} = \int \phi^*_{\sigma,k}(x)\hat{\Psi}_\sigma(x) dx,
\end{equation}
where $\hat{\Psi}_\sigma(x) = \sum\limits_k \phi_{\sigma,k}(x)\hat{a}_{\sigma,k} $ denotes the bosonic field operator that annihilates a $\sigma$-species boson in the single-particle state $\phi_{\sigma,k}(x)$ at position $x$. As usual, the annihilation operator $\hat{a}_{\sigma,k}$ and its corresponding creation operator $\hat{a}^\dagger_{\sigma,\ell}$ must satisfy the commutation relations
\begin{align}
    \left[\hat{a}_{\sigma,k},\hat{a}_{\sigma^\prime,\ell}^\dagger \right] &= \delta_{\sigma\sigma^\prime}\delta_{k\ell}, \\ 
	\left[\hat{a}_{\sigma,k}^\dagger,\hat{a}_{\sigma^\prime,\ell}^{\dagger}\right]  &=\left[\hat{a}_{\sigma,k},\hat{a}_{\sigma^\prime,\ell}\right] = 0.
\end{align}
The many-body Hamiltonian can then be transformed into 
\begin{align}
	\label{full_hamiltonian_second}
	\hat{H} = & \sum_{\sigma\in\{A,B,C\}}\left[\sum_{k,\ell}  h^{\sigma}_{k\ell} \hat{a}^\dagger_{\sigma,k} \hat{a}^{}_{\sigma,\ell} + \dfrac{1}{2}\sum_{k\ell mn}  W^{\sigma}_{k\ell mn} \hat{a}^\dagger_{\sigma,k} \hat{a}^\dagger_{\sigma,\ell} \hat{a}^{}_{\sigma,m}\hat{a}^{}_{\sigma,n} \right]+ \sum_{\sigma\neq\delta\in\{A,B,C\}}\sum_{k\ell m n}W^{\sigma\delta}_{k\ell mn} \hat{a}^\dagger_{\sigma,k} \hat{a}^\dagger_{\delta,\ell} \hat{a}^{}_{\sigma,m} \hat{a}^{}_{\delta,n},
\end{align}
where $h^{\sigma}_{k\ell}$ denotes the $\sigma$-component one-body matrix elements, while $W^{\sigma}_{k\ell mn}$, and $W^{\sigma\delta}_{k\ell mn}$ are the intra- and interspecies two-body matrix elements, respectively. In this work, we use the harmonic oscillator eigenfunctions as the single-particle functions $\phi_{\sigma,k}(x)$ since this choice not only makes $h^{\sigma}_{k\ell} = \left(k +\dfrac{1}{2}\right)\delta_{k\ell}$ diagonal, \TT{but also allows us to employ the effective-interaction approach for obtaining the matrix elements $W^{\sigma}_{k\ell mn} $ and $W^{\sigma\delta}_{k\ell mn}$ from the analytic solution of the two-body problem \cite{busch1998two}. This regularized strategy has been widely used in previous works and it has been thoroughly shown that it gives a better convergence in the case of the Fermi-Huang $\delta$-function pseudo-potential~\cite{rotureau2013interaction,lindgren2014fermionization,dehkharghani2015quantum,anhtai2023quantum,rammelmuller2023modular}. We remark that our effective-interaction approach is applicable solely to the parabolic trap, while a recently established regularized scheme incorporating the full two-body spectrum can be used for any trapping potentials~\cite{brauneis2024comparison}.} 

We \TT{next} solve the many-body Hamiltonian by expanding the trial wavefunction (ansatz) into a linear combination of a set of orthonormal Fock states associated with the expansion coefficients $c_{j_{\rm{A}},j_{\rm{B}},j_{\rm{C}}}$ as
\begin{align}
    \label{eq:ansatze}
	|\Psi\rangle = \sum_{j_{\rm{A}}=1}^{D_\text{A}} \sum_{j_{\rm{B}}=1}^{D_\text{B}}\sum_{j_{\rm{C}}=1}^{D_\text{C}}c_{j_{\rm{A}},j_{\rm{B}},j_{\rm{C}}} |n^\text{A}\rangle_{j_{\rm{A}}} |n^\text{B}\rangle_{j_{\rm{B}}} |n^\text{C}\rangle_{j_{\rm{C}}} = \sum_{J=1}^{D} c_J|J\rangle,
\end{align}
\TT{where $|n^{\sigma}\rangle_{j_\sigma} = |n^\sigma_1, n^\sigma_2, \dots, n^\sigma_k, \dots, n^\sigma_{M_\sigma}\rangle$ denotes the $j_\sigma$-th permanent of species $\sigma$ that characterizes a configuration of $N_\sigma$ bosons distributed over $M_\sigma$ single-particle functions. The occupation numbers, $n^\sigma_k$, can be positive integers varying between 0 and $N_\sigma$ and satisfy the condition $\sum\limits_{k=1}^{M_\sigma} n_k^\sigma = N_\sigma$. }For brevity, we use a composite index such that $c_J = c_{j_{\rm{A}},j_{\rm{B}},j_{\rm{C}}}$ and $|J\rangle = |n^\text{A}\rangle_{j_{\rm{A}}} |n^\text{B}\rangle_{j_{\rm{B}}} |n^\text{C}\rangle_{j_{\rm{C}}}$. For numerical reasons the many-body Fock basis has to be truncated such that a sufficiently large but finite Hilbert space is used and the total number of Fock states in the expansion is $D = D_\text{A}\cdot D_\text{B}\cdot D_\text{C}$ with $D_\sigma$ being the number of $\sigma$-species Fock states. In this work we use an efficient truncation scheme proposed in Refs.~\cite{chrostowski2019efficient,plodzien2018numerically} to determine the value of $D$, which limits the many-body Fock states $|J\rangle$ in the expansion to ones that have an energy smaller than a certain optimal value $E_{opt}$. This allows one to control the accuracy of the numerical results by varying $E_{opt}$. This technique can be applied to both bosonic and fermionic systems and has been widely used in recent years \cite{anhtai2023quantum,Tai2024inprepPRX,rammelmuller2023modular,becker2024synthetic,pkecak2020signatures,rojo2022direct,lydzba2022signatures,rojo2024few}. To make the calculations in the present work feasible, we use another technique that significantly reduces the dimension of the truncated Hilbert by selecting dominant configurations with respect to the spatial symmetry of the desired many-body state \cite{anhtai2023quantum}. If the trapping potential is spatially symmetric 
\begin{equation}
    V(x) = V(-x),
\end{equation}
the single-particle eigenfunctions $\phi_n(x)$ have a well-defined parity given by 
\begin{equation}
    \hat{P} \phi_n(x) = p\phi_n(-x),
\end{equation}
where $\hat{P}$ is the symmetry operator whose eigenvalues are $p= \pm 1$. The single-particle functions $\phi_n(x)$ with $p=1$ are spatially symmetric or even functions, while those with $p=-1$ are spatially antisymmetric or odd functions. Since the Fock states are constructed as the symmetrized Hartree product of the single-particle functions, they also satisfy this spatial symmetry. This allows us to classify the Fock states into two categories according to their spatial symmetries: even- and odd-parity Fock states. As a consequence, the many-body wave functions only span in one of these subspaces. Therefore, in practice, an ansatz can be constructed only from Fock states that have the same parity as the desired many-body wavefunction. \TTT{Since the harmonic trap is spatially symmetric, the many-body Hamiltonian given by Eq.~\eqref{eq:full_hamiltonian_first} is invariant under the transformation $x_i^\sigma \to -x_i^\sigma$. The ground states of the multi-species few-boson systems are therefore spatially symmetric, which justifies to expand the ansatz in Eq.~\eqref{eq:ansatze} using only Fock states $|J\rangle$ that have even-parity symmetry.}

Since we focus on studying the stationary properties, the problem of variationally finding the expansion coefficients $c_{J}$ such that the expectation value of the Hamiltonian \eqref{full_hamiltonian_second} is minimized with respect to the ansatz \eqref{eq:ansatze} now leads to a standard Hermitian eigenvalue problem which can be written as
\begin{equation}
	\label{manybody_equations}
	\mathbf{\mathcal{H}}|C_m\rangle = E_m|C_m\rangle,
\end{equation}
where $\mathbf{\mathcal{H}} = \langle I |\hat{H} |J \rangle $ is the matrix representation of the many-body Hamiltonian \eqref{full_hamiltonian_second}. The $m$-th eigenvalue is given by $E_m$ and the corresponding eigenvector $|C_m\rangle$ is a column vector storing the expansion coefficients $c_{J}$. So far this improved scheme has been employed in our previous works \cite{Tai2024inprepPRX,anhtai2023quantum}. \TTT{The details of the numerical setup and convergence of our ab initio study are presented in Appendix.~\ref{appendixx}.}

\subsection{Quantities of interest}
Having obtained the full many-body wavefunction for a given set of parameters, we are able to investigate all static properties of the ground-state. \TT{As mentioned before, we are interested in the quantum correlations and entanglement in the system, specifically the one- and two-body correlations. To quantify the inter- and intraspecies correlations present in our system, including correlations that arise from direct interactions between particles from the same species and correlations that are induced by the couplings to other species, we evaluate the inter- and intraspecies bipartite mutual information (BMI) respectively defined as 
\begin{align}
    I_{\sigma\delta} & = S_{\sigma\sigma} + S_{\delta\delta} - S_{\gamma\gamma}, \\ 
    I_{\sigma}       & = 2S_{\sigma} - S_{\sigma\sigma}.
\end{align}
Here $S_{\sigma} = -\text{tr}\left[\tilde{\rho}_{\sigma}\log_2\left(\Tilde{\rho}_{\sigma}\right)\right]$ and $S_{\sigma\delta} = -\text{tr}\left[\Tilde{\rho}_{\sigma\delta}\log_2\left(\Tilde{\rho}_{\sigma\delta}\right)\right]$ are the single- and two-particle von Neumann entropies and the BMI is always non-negative \cite{amico2008entanglement}. The matrix elements of the reduced one-body density matrix of one $\sigma$-species boson, $\Tilde{\rho}_{\sigma}$, are given by 
\begin{equation}
    \left(\Tilde{\rho}_{\sigma}\right)_{k\ell} = \langle\Psi |\hat{a}^\dagger_{\sigma,k} \hat{a}_{\sigma,\ell}  |\Psi\rangle,
\end{equation}
with $|\Psi\rangle$ being the ground-state wavefunction of the system. Meanwhile, the $\Tilde{\rho}_{\sigma\sigma^\prime}$ denote the reduced two-body density matrix of one $\sigma$-species and one $\sigma^\prime$-species boson whose elements are defined as
\begin{equation}
    \left(\Tilde{\rho}_{\sigma\sigma^\prime}\right)_{k\ell mn} = \langle\Psi |\hat{a}^\dagger_{\sigma,k} \hat{a}^\dagger_{\sigma^\prime,\ell}\hat{a}_{\sigma,m}\hat{a}_{\sigma^\prime,n}|\Psi\rangle.
\end{equation}
Note that $\sigma^\prime$ can either coincide with or differ from $\sigma$. We further analyze the different density matrices by calculating their corresponding two-point correlation functions in spatial coordinates. For the reduced one-body density matrix (OBDM), which describes \TT{the one-body coherence between the two points $x$ and $x^\prime$}, it is given by
\begin{align}
\label{eq:OBDM}
\rho^{(1)}_{\sigma}(x,x^\prime) = \sum_{k,\ell} \left(\Tilde{\rho}_{\sigma}\right)_{k\ell} \phi^*_{\sigma,k}(x^\prime) \phi_{\sigma,\ell}(x).
\end{align}
The ascendingly sorted eigenvalues $\lambda^\sigma_j$ of the OBDM describe the occupations of the corresponding natural orbitals and characterize the coherence/fragmentation according to Penrose-Onsager criterion \cite{penrose1956Bose}. Furthermore, the diagonal of the OBDM defines the one-body density distribution  
\begin{equation}
    \label{eq:OBDF}
    \rho^{(1)}_{\sigma}(x)= \sum_{k,\ell} \left(\Tilde{\rho}_{\sigma}\right)_{k\ell} \phi^*_{\sigma,k}(x) \phi_{\sigma,\ell}(x),
\end{equation}
which can be used to assess the spatial distribution of the three individual components. Spatial correlations can be further characterized by the intra- and interspecies two-body correlation functions (TBCF) which are respectively defined as
\begin{align}
\label{eq:TBCF}
\rho^{(2)}_{\sigma}(x_1,x_2) &= \sum_{k,\ell,m,n} \left(\Tilde{\rho}_{\sigma\sigma}\right)_{k\ell mn} \phi^*_{\sigma,k}(x_1) \phi^*_{\sigma,\ell}(x_2)\phi_{\sigma,m}(x_1) \phi_{\sigma,n}(x_2)\\
\rho^{(2)}_{\sigma\delta}(x^{\sigma},x^{\delta}) &= \sum_{k,\ell,m,n} \left(\Tilde{\rho}_{\sigma\delta}\right)_{k\ell mn}\phi^*_{\sigma,k}(x^\sigma) \phi^*_{\delta,\ell}(x^\delta)\phi_{\sigma,m}(x^\sigma) \phi_{\delta,n}(x^\delta)
\end{align}
The physical meaning of $\rho^{(2)}_{\sigma}(x_1,x_2)$ is the joint probability of finding one $\sigma$-species boson at position $x_1$ and the other of the same species at $x_2$. Similarly, $\rho^{(2)}_{\sigma\delta}(x^{\sigma},x^{\delta})$ has the same interpretation as $\rho^{(2)}_{\sigma}(x_1,x_2)$, but for two bosons of different species. It is worth noting that these spatial correlations functions can be experimentally measured via the time-of-flight absorption imaging technique \cite{bergschneider2019experimental,torma2014quantum,guo2024observation}. To maintain consistency we will in the following normalize all density profiles to each component's respective particle number and the trace of all spatial density matrices to unity. }

\section{Results and Discussion}
\label{sec:discussion}
\subsection{Tri-correlated states}
\TT{In the following we focus on three representative phases that are labeled in Fig.~\ref{fig:ice-cube}.} We will show that these phases possess interesting long-range correlations or spatial self-localization due to the presence of the third species. The remaining tri-correlated states are presented in Appendix~\ref{appendix}. \TT{It is worth noting again that in the following we present the results for the minimal three-species system that has $N_\sigma=2$ bosons in each species. }

\subsubsection{Fermionized Phase Separation}
\label{FPS}
\begin{figure*}[tb]
    \centering
     \includegraphics[width=\textwidth]{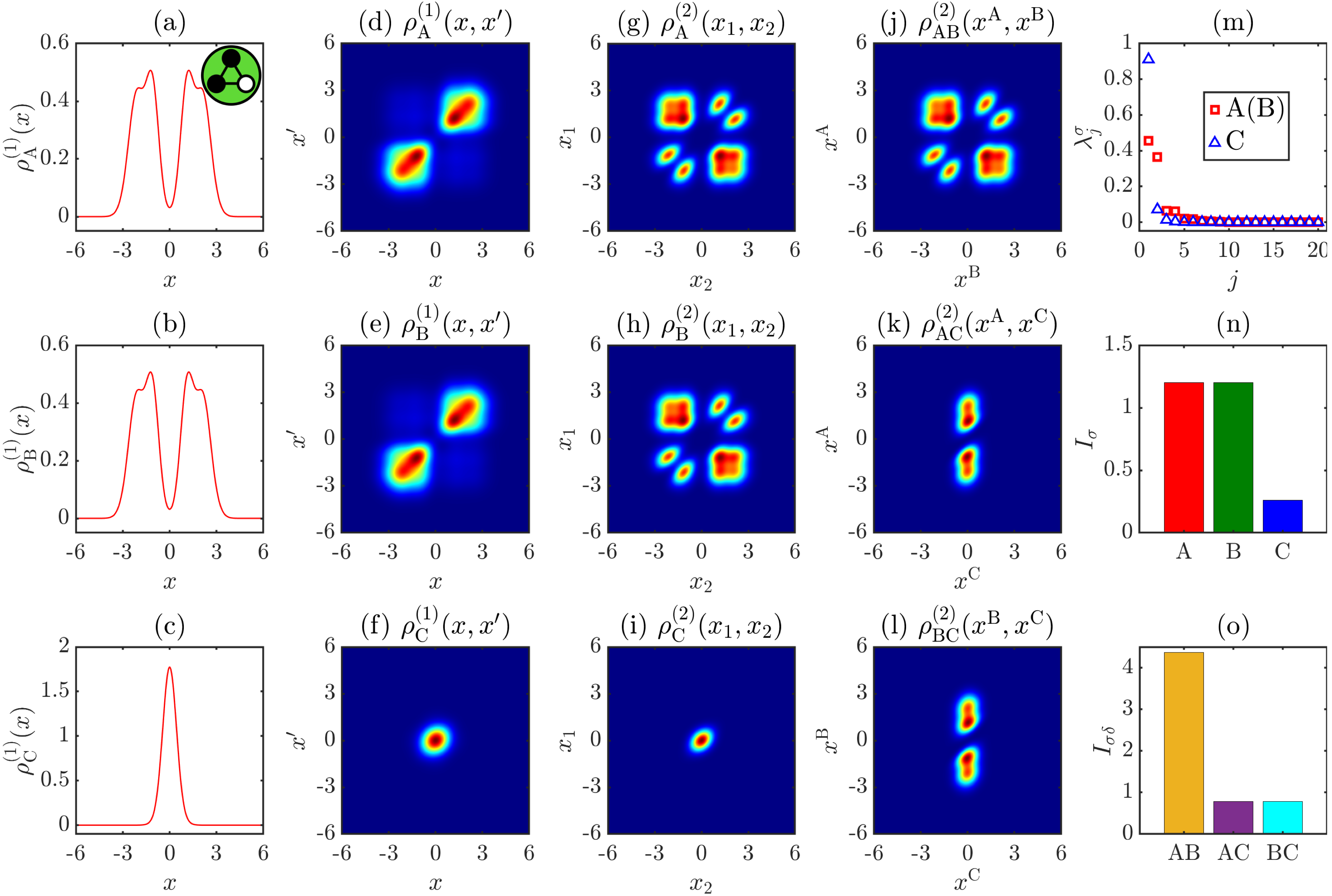}
    \caption{The ``Fermionized Phase Separation" phase ($g_{\rm{A}}=g_{\rm{B}}=g_{\rm{AB}}=g_{\rm{BC}}=g_{\rm{AC}}\to\infty$, $g_{\rm{C}}=0$). (a-c) The one-body density distribution function $\rho_\sigma^{(1)}(x)$. (d-f) The one-body density matrix $\rho_\sigma^{(1)}(x,x^\prime)$. (g-i) The intraspecies two-body correlation function $\rho^{(2)}_{\sigma}(x_1,x_2)$. (j-l) The interspecies two-body correlation function $\rho^{(2)}_{\sigma\delta}(x^{\sigma},x^{\delta})$. (m) The eigenvalues $\lambda^\sigma_j$ of the one-body density matrices. (n) The intraspecies mutual information $I_\sigma$. (o) The interspecies mutual information $I_{\sigma\delta}$. \TT{In panels (d)-(l), the color gradient ranges from minimum (blue) to maximum (dark red).}}
    \label{fig:fermionized_phase_separation}
\end{figure*}

\TT{The first phase we discuss has isotropic interspecies interactions $g_{\rm{AB}}=g_{\rm{BC}}=g_{\rm{AC}}\to\infty$, and intraspecies interactions $g_{\rm{A}}=g_{\rm{B}}\rightarrow\infty$ and $g_{\rm{C}}=0$. This state is invariant under exchange of species A and B and is labeled \textbf{1} in Fig.~\ref{fig:ice-cube}(a)}. \TT{We term this the ``Fermionized Phase Separation" phase due to the properties of the densities and correlation functions shown in Fig.~\ref{fig:fermionized_phase_separation}}. \TT{Let us first discuss} the ground-state density profiles as shown in Figs.~\ref{fig:fermionized_phase_separation}(a-c). One can immediately see that the C species occupies the center of the trap, while the A and B species are spatially separated to the left and right edges of the trap. This is very similar to the phase separation case in binary bosonic mixtures~\cite{garcia2014quantum}, and can be straightforwardly understood by realizing that the interaction energy is minimized when the species with zero intraspecies interactions is localized in the high-density trap center, while the species with repulsive intraspecies interactions achieve lower densities by splitting and reducing the overlap of the two particles. 

However, this case distinguishes itself from the phase separation case in binary mixtures as the overlapping parts of A and B are fermionized. To fully understand the behavior of the two A and two B atoms in this case, it is necessary to examine their OBDMs (Figs.~\ref{fig:fermionized_phase_separation}(d-f)) and TBCFs (Figs.~\ref{fig:fermionized_phase_separation}(g-l)). While the OBDMs clearly show the splitting of the A and B components, the TBCFs show how the particles in each component are organized. For instance, for the A component there are both anti-correlated contributions, where the particles of A are split on either side of C, and correlated contributions, where both particles of A are bunched together on one side of C. For the latter the effects of the strong intraspecies interactions are clearly seen in the vanishing of the diagonal contribution. The TBCF for the B particles is exactly the same, and so is the interspecies TBCF $\rho^{(2)}_\text{AB}$, showing that the A and B particles sit on top of each other in a fully fermionized state. We note that since all interaction strengths are the same, the A-B system is SU(2) symmetric and thus any A particle can be swapped with a B particle. Therefore, any distribution of two particles on each side of C would have the same energy \TT{and thus the groundstate is doubly degenerate.}

To understand the coherence and correlation properties of the ground state we look at the eigenvalues of the OBDMs shown in Fig.~\ref{fig:fermionized_phase_separation}(m). The A(B) species can be seen to be fragmented with the eigenvalues being nearly doubly degenerate due to the spatial splitting into a superposition state between the left and right hand side of the harmonic trap. This indicates that the A and B species both possess strong intra- and interspecies correlations that can be quantified by their high mutual information $I_\text{A(B)}$ and $I_\text{AB}$ shown in Figs.~\ref{fig:fermionized_phase_separation}(n,o). Meanwhile, the C component remains mostly coherent with one dominant eigenvalue $\lambda_1^\text{C} \approx 0.9$. However, a second relevant eigenvalue $\lambda_2^\text{C} \approx 0.1$ is also visible, despite the absence of an intraspecies interaction between C-type bosons. This is consistent with the mutual information $I_\text{C}$ also having a finite value, indicating that the two C atoms are actually correlated (see Fig.~\ref{fig:fermionized_phase_separation}(n)), which stems from the induced effective attractive interactions through the interspecies couplings to the A and B species which weakly binds the C particles as shown in Figs.~\ref{fig:fermionized_phase_separation}(f,i). Finally, one can see that the C species is less correlated with the A(B) species as there is reduced overlap between the states due to phase separation and therefore the interspecies mutual information takes comparatively small values. 

\subsubsection{Correlation-induced Anti-bunching}
\label{CIABI}
\begin{figure*}[tb]
    \centering
    \includegraphics[width=\textwidth]{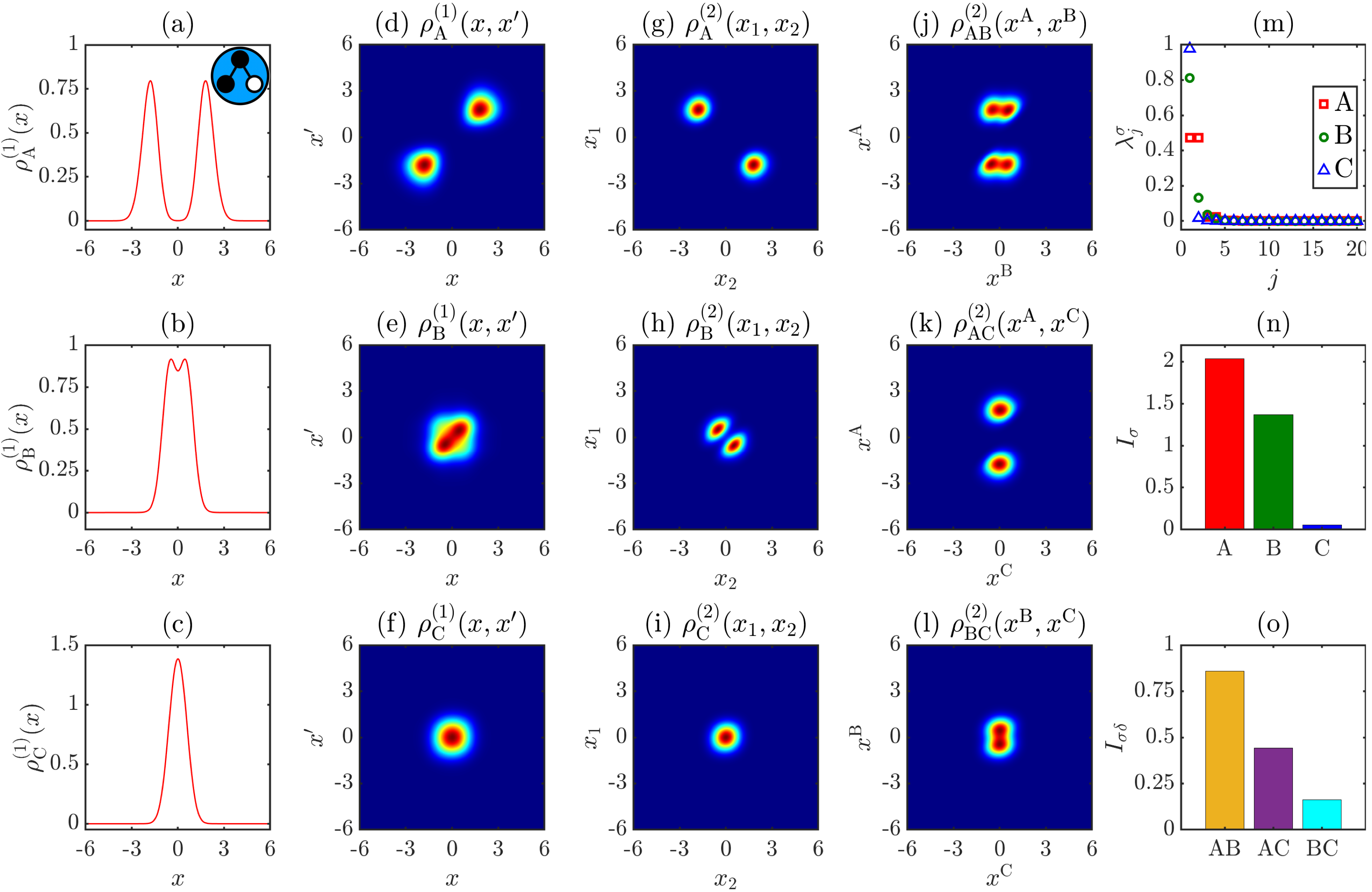}
    \caption{The `Correlation-induced Anti-bunching' phase ($g_{\rm{A}}=g_{\rm{B}}=g_{\rm{AB}}=g_{\rm{AC}}\to\infty$, and $g_{\rm{C}}=g_{\rm{BC}}=0$). (a-c) The one-body density distribution function $\rho_\sigma^{(1)}(x)$. (d-f) The one-body density matrix $\rho_\sigma^{(1)}(x,x^\prime)$. (g-i) The intraspecies two-body correlation function $\rho^{(2)}_{\sigma}(x_1,x_2)$. (j-l) The interspecies two-body correlation function $\rho^{(2)}_{\sigma\delta}(x^{\sigma},x^{\delta})$. (m) The eigenvalues $\lambda^\sigma_j$ of the one-body density matrices. (n) The intraspecies mutual information $I_\sigma$. (o) The interspecies mutual information $I_{\sigma\delta}$. \TT{In panels (d)-(l), the color gradient ranges from minimum (blue) to maximum (dark red).}}
    \label{fig:phase_separation_squeezed_TG}
\end{figure*}

\TT{Next we discuss a phase which has anisotropic interspecies interactions, $g_{\rm{AB}}=g_{\rm{AC}}\to\infty$ and $g_{\rm{BC}}=0$, while the intraspecies interactions are $g_{\rm{A}}=g_{\rm{B}}\rightarrow\infty$ and $g_{\rm{C}}=0$ and which is labeled \textbf{2} \TT{in Fig.~\ref{fig:ice-cube}(b)}. This system does not possess any invariant particle exchanges and therefore possesses a more complex distribution of the particles and their correlations which we show in Fig.~\ref{fig:phase_separation_squeezed_TG}.} While for this set of interaction strengths the TG pairs of A and B atoms would demonstrate full fermionization in the absence of the third species, the different interactions with the C species leads to a completely different behaviour. From Figs.~\ref{fig:phase_separation_squeezed_TG}(a-c) one can immediately see that the species B and C locate in the center of the trap, while species A is anti-bunched due to its strong repulsive intraspecies interactions and located at the edges with one atom on each side of the central clouds (see Fig.~\ref{fig:phase_separation_squeezed_TG}(g)). \TT{Based on these observations we call this the ``Correlation-induced Anti-bunching" phase.} The spatial superposition state formed by species A can also be seen in the doubly-degenerate natural occupation numbers, in which the two largest values are close to $0.5$ \cite{murphy2007boson}, while the large value of $I_\text{A}$ in Fig.~\ref{fig:phase_separation_squeezed_TG}(n) shows strong intraspecies correlations as expected. \TTT{Meanwhile the numerical value of the largest natural occupancy for the C species is close to one, which means that it remains mostly coherent. This is consistent with the Gaussian-like shape of the correlation functions depicted in Figs.~\ref{fig:phase_separation_squeezed_TG}(f,i), which, in a harmonic trap, would be purely Gaussian for separable states. Interestingly, the shape of $\rho^{(1)}_\text{C}(x,x^\prime)$ is in fact modified by the weak induced interspecies correlations via the strongly repulsive coupling to species A as can be seen from $I_\text{C}$ slightly deviating from zero in Fig.~\ref{fig:phase_separation_squeezed_TG}(n).}


The most remarkable self-organization effect of this case is that species B is located in the center of the trap despite its strongly repulsive intraspecies interaction. While it forms a localized TG state, one can see in Fig.~\ref{fig:phase_separation_squeezed_TG}(b) that the width of $\rho^{(1)}_\text{B}(x)$ is noticeably smaller than that of the conventional TG pair in a harmonic trap \cite{girardeau2001ground}, seemingly due to the repulsive pressure from the A species atoms. This is also confirmed by the one- and two-body correlation functions shown in Figs.~\ref{fig:phase_separation_squeezed_TG}(e,h). However, we point out that the correlations between the B particles are noticeably reduced when compared to two TG particles in a harmonic oscillator (see Fig.~\ref{fig:phase_separation_squeezed_TG}(n)) in which case $I_\text{B}\approx1.97$ \cite{murphy2007boson}, as the large coupling to the A component is responsible for screening the correlations between the B particles. Indeed, the A and B components are strongly correlated, as seen in Fig.~\ref{fig:phase_separation_squeezed_TG}(o), which, due to entanglement monogamy, reduces the correlations in B \cite{coffman2000distributed}. Finally, the correlations between B and C are non-zero even though they are not directly coupled ($g_\text{BC}=0$), which is again due to induced correlations from their mutual coupling to the A species. Finally, we note that in appendix~\ref{CIABII} we discuss a similar phase in which the intraspecies interactions in C has been changed to $g_{\rm{C}}\rightarrow\infty$. 

\subsubsection{Correlation-induced Bunching}
\label{CIBII}
\begin{figure*}[tb] 
    \centering
    \includegraphics[width=\textwidth]{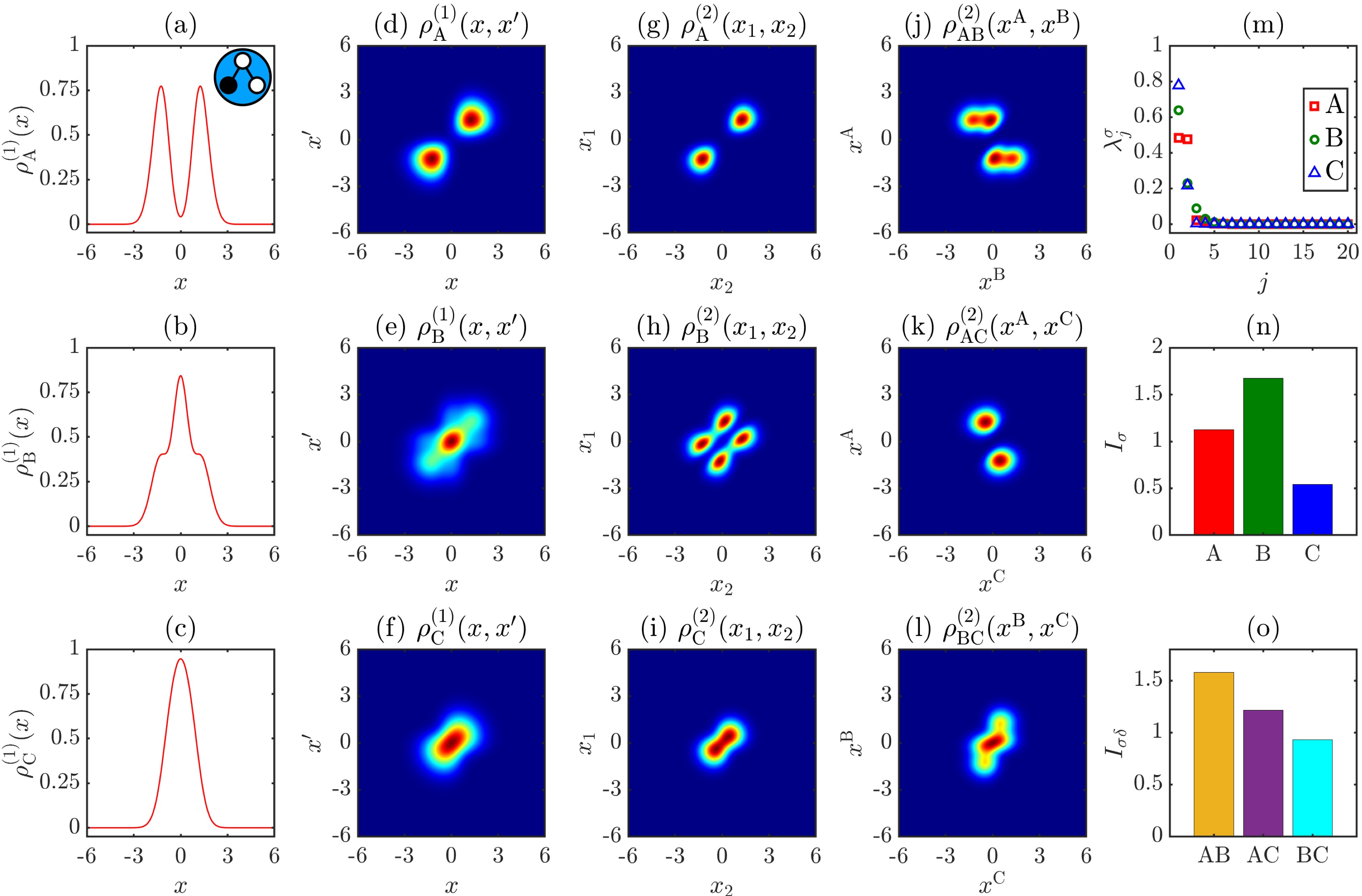}
    \caption{The ``Correlation-induced Bunching Type II" phase ($g_{\rm{AB}}=g_{\rm{AC}}=g_{\rm{B}}\to\infty$, and $g_{\rm{A}}=g_{\rm{C}}=g_{\rm{BC}}=0$). (a-c) The one-body density distribution function $\rho_\sigma^{(1)}(x)$. (d-f) The one-body density matrix $\rho_\sigma^{(1)}(x,x^\prime)$. (g-i) The intraspecies two-body correlation function $\rho^{(2)}_{\sigma}(x_1,x_2)$. (j-l) The interspecies two-body correlation function $\rho^{(2)}_{\sigma\delta}(x^{\sigma},x^{\delta})$. (m) The eigenvalues $\lambda^\sigma_j$ of the one-body density matrices. (n) The intraspecies mutual information $I_\sigma$. (o) The interspecies mutual information $I_{\sigma\delta}$. \TT{In panels (d)-(l), the color gradient ranges from minimum (blue) to maximum (dark red).}}
    \label{fig:cib_phase_separation_type_II}
\end{figure*}

While the two phases discussed above both show splitting between infinitely repulsive interacting bosons, we next discuss a phase that is dominated by completely different physics stemming from the presence of induced attractive interactions, that can lead to stronger localisation of non-interacting bosons. \TT{Again we consider anisotropic interspecies interactions, $g_{\rm{AB}}=g_{\rm{AC}}\to\infty$ and $g_{\rm{BC}}=0$, while only the B component has strong intraspecies interactions so $g_{\rm{A}}=g_{\rm{C}}=0$ and $g_{\rm{B}}\rightarrow\infty$. This phase is labeled \textbf{3} in Fig.~\ref{fig:ice-cube}(b)} \TTT{and again possesses no species exchange symmetry.} One can see in Figs.~\ref{fig:cib_phase_separation_type_II}(a-c) that this leads to a situation where species B exhibits a unique density profile due to these competing interactions, possessing a Gaussian-like peak around $x=0$ which noticeably widens around the half maximum into distinct shoulders. This shape is the consequence of the competition between the pressure from the A component to phase separate and the intraspecies interaction to expand to allow the B atoms to decrease their overlap with each other. \TT{For this reason we call this phase ``Correlation-induced Bunching''}.

The TBCF of species B, $\rho^{(2)}_\text{B}(x_1,x_2)$, shows a zero along the diagonal due to the fact that the two B atoms cannot be in the same place simultaneously, but also highlights a non-trivial ordering of the B particles. If one B particle is localized at the trap minimum $x=0$, the other B particle will be localized in a superposition of being to the left and right of it. This \TT{therefore} leads to the fact that the B bosons are more strongly correlated than A bosons, which only possess bunching correlations. Indeed, components A and C behave similarly to those in the ``Correlation-induced Bunching type II" case described in the appendix~\ref{CIBI}. In Fig.~\ref{fig:cib_phase_separation_type_II}(o) we can see $I_\text{AB}>I_\text{AC}>I_\text{BC}$, which shows that the components A and B have the highest interspecies correlations due to the increased overlap, while the components B and C have the lowest interspecies correlations as they are induced only. \TT{Note that if the C species also has infinite intraspecies interactions, i.e.~$g_{\rm{AB}}=g_{\rm{AC}}=g_{\rm{C}}=g_{\rm{B}}\to\infty$, and $g_{\rm{A}}=g_{\rm{BC}}=0$, the system exhibits similar correlations and this phase is discussed in Appendix~\ref{CIBIII}.} 

\subsection{Crossover between phases}
\label{subsec:phase_transition}
\begin{figure*}[tb]
    \centering
    \includegraphics[width=\textwidth]{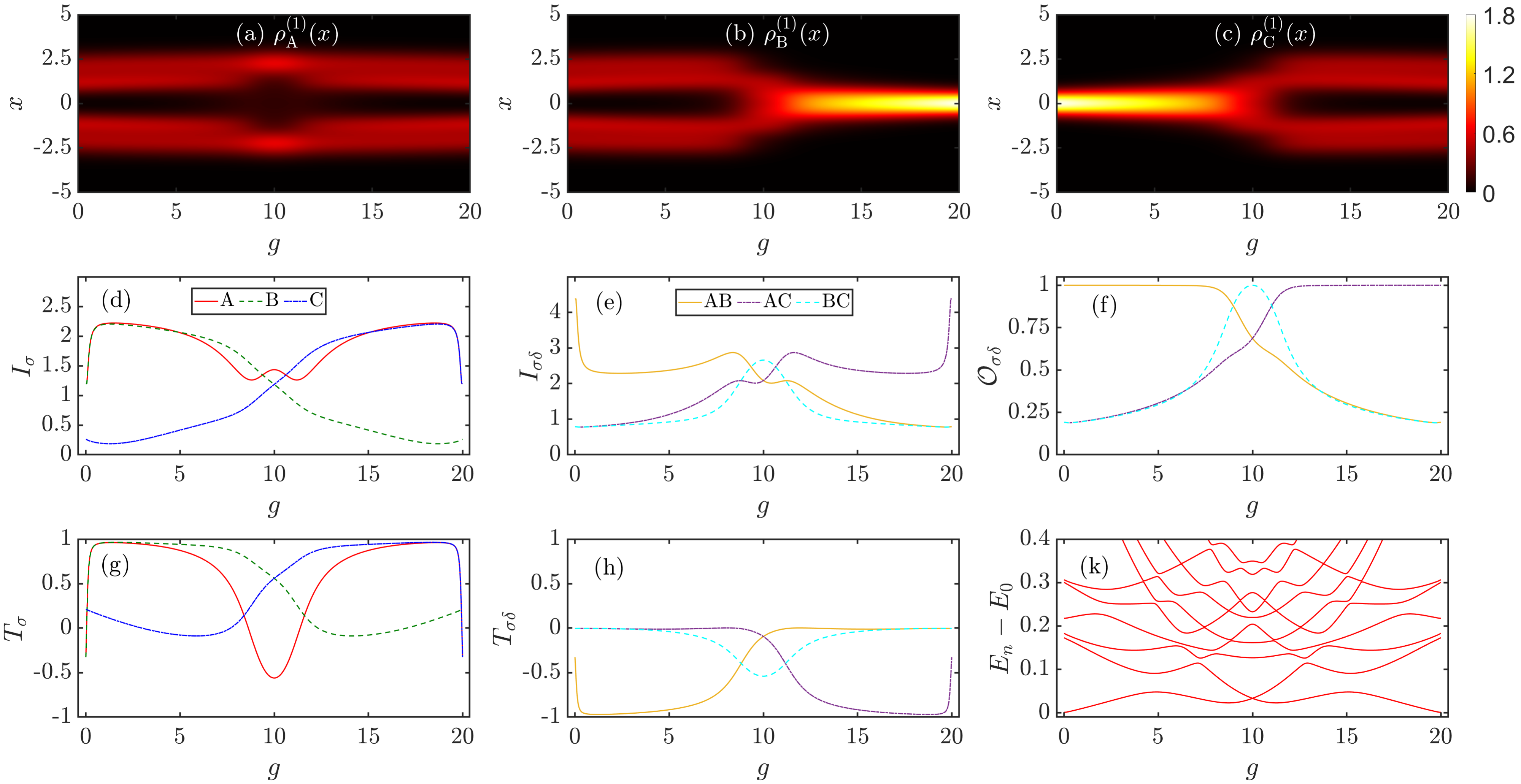}
    \caption{\TT{The crossover between between the Fermionized Phase Separation phase from the initial state $g_{\rm{A}}=g_{\rm{B}}=g_{\rm{AB}}=g_{\rm{AC}}=g_{\rm{BC}}\to\infty,g_{\rm{C}}=0$ to the final state $g_{\rm{A}}=g_{\rm{C}}=g_{\rm{AB}}=g_{\rm{AC}}=g_{\rm{BC}}\to\infty,g_{\rm{B}}=0$}. (a-c) The spatial one-body density distribution function $\rho_\sigma^{(1)}(x)$. (d) The intraspecies mutual information $I_\sigma$. (e) The interspecies mutual information $I_{\sigma\delta}$. (f) The density overlap $\mathcal{O}_{\sigma\delta}$. (g) The intraspecies correlation crossover $T_\sigma$. (h) The interspecies correlation crossover $T_{\sigma\delta}$. (k) The even-parity excitation energy spectrum $E_n-E_0$. \TT{Note that $g_\text{C}=g$ while $g_\text{B}=20-g$.}}
    \label{fig:phase_transition_1}
\end{figure*}

\TT{While the vertices of the cubes in Fig.~\ref{fig:ice-cube} most clearly indicate the different possible phases, there is a large state space in between where the interaction strengths can be finite. The crossover region between two vertices will therefore be non-trivial, particularly if the limits are strongly correlated states. As a representative example we therefore look at the Fermionized Phase Separation phase and consider the direct connection between vertices \textbf{1} and \textbf{4} in Fig.~\ref{fig:ice-cube}(a).} 
\TTT{These two states are related through reflection symmetry and we explore their crossover by} increasing $g_\text{C}=0\rightarrow\infty$ while simultaneously decreasing $g_\text{B}=\infty\rightarrow 0$. Along this trajectory strong interspecies correlations will be swapped from the A-B pair to the A-C pair. Since numerically we only treat the strong interactions up to $g=20$, we show in Fig.~\ref{fig:phase_transition_1} the corresponding results as a function of increasing $g_\text{C}=g=0\rightarrow20$, while $g_\text{B}=20-g_\text{C}$. The spatial one-body density distribution functions $\rho_\sigma^{(1)}(x)$ are shown in the first row of Fig.~\ref{fig:phase_transition_1}. As already discussed in Fig.~\ref{fig:fermionized_phase_separation} both species A and B are in a phase-separated fully fermionized state and are located at the edges of the trap for $g=0$, while species C is tightly localized in the center. Once the coupling strength $g$ increases \TT{the degeneracy of the ground state is broken (see the energy spectrum in Fig.~\ref{fig:phase_transition_1}(k))}, \TT{however the density of species A and B  completely overlap until $g\approx 8$}. This can be quantified by the overlap between the one-body distribution of two species, $\rho^{(1)}_\sigma(x)$ and $\rho^{(1)}_\delta(x)$, which is given by 
\begin{equation}
    \mathcal{O}_{\sigma\delta} = \Big|\int \sqrt{\rho^{(1)}_\sigma(x) \rho^{(1)}_\delta(x)} dx\Big|^2\,.
\end{equation}
This will be unity if two species are exactly superimposed, as shown for the A and B species in Fig.~\ref{fig:phase_transition_1}(f) for $g\lesssim 8$. For larger interactions there is a crossover region where the B particles swap positions with the C particles, and for interactions $g\gtrsim 12$ the A and C species have maximum overlap. 

While the density and its overlap can give some idea of the re-organization of the particles, it contains no information about the position of particles with respective to one another, i.e.~whether they are bunched or anti-bunched as described by the TBCF. The crossover between anti-bunching and bunching correlations in species $\sigma$ can be well quantified by the intraspecies two-particle coincidence function
\begin{equation}
    T_\sigma = \iint\limits_{x_1.x_2>0} \rho^{(2)}_\sigma(x_1,x_2)dx_1dx_2 - \iint\limits_{x_1.x_2<0}\rho^{(2)}_\sigma(x_1,x_2)dx_1dx_2
\end{equation}
which compares the probability of finding two $\sigma$-type particles being bunched (the first integral) with being anti-bunched (the second integral). It should be remarked that the quadrant defined by the condition $x_1.x_2>0$ encompasses two spatial regions where the variables $x_1$ and $x_2$ have the same sign, specifically $(x_1 <0, x_2<0)$ and $(x_1 >0, x_2>0)$, illustrating bunching correlations. Meanwhile, the quadrant \TT{that} satisfies the condition $x_1.x_2<0$ corresponds to the area where the variables $x_1$ and $x_2$ have the opposite sign, showing anti-bunching correlations. If $T_\sigma>0$ ($T_\sigma<0$) the bunching (anti-bunching) correlations are more dominant than the anti-bunching (bunching) ones, while two $\sigma$-species particles are said to be fully bunched if $T_\sigma=1$, and fully anti-bunched if $T_\sigma=-1$. In Fig.~\ref{fig:phase_transition_1}(g) we show the intraspecies two-particle coincidence function as a function of $g$. The initial state at $g=0$ is slightly more anti-bunched in the A and B components (see Fig.~\ref{fig:fermionized_phase_separation}), however they become maximally bunched for small and finite $g>0$. In this case when the symmetry between the A and B species is slightly broken ($g_A\neq g_B$) particles of the same species will more likely stay together, but still bisected by the C component. Interestingly, the opposite effect is seen in the interspecies two-particle coincidence function, which can similarly characterized by
\begin{equation}
    T_{\sigma\delta} = \iint\limits_{x^\sigma.x^\delta>0} \rho^{(2)}_{\sigma\delta}(x^\sigma,x^\delta)dx^\sigma dx^\delta - \iint\limits_{x^\sigma.x^\delta<0} \rho^{(2)}_{\sigma\delta}(x^\sigma,x^\delta)dx^\sigma dx^\delta.
\end{equation}
This function is shown in Fig.~\ref{fig:phase_transition_1}(h) and it is similarly slightly anti-bunched at $g=0$ for an A-B pair, since the inter- and intraspecies TBCFs are identical in this case (see Fig.~\ref{fig:fermionized_phase_separation}(g,j,h)). When the the symmetry is broken the A and B species maximally anti-bunch with respect to one another, i.e.~if an A particle is found on the left-side of the trap, a B particle will be found on the right-side of the trap. The tendency of the particles to bunch or anti-bunch is also echoed in the intra- and interspecies mutual information as shown in Fig.~\ref{fig:phase_transition_1}(d) and Fig.~\ref{fig:phase_transition_1}(e) respectively. For example, the bunching of A particles leads to an increase in their intraspecies mutual information, while the anti-bunching between A and B particles reduces their interspecies mutual information.

\begin{figure*}[tb]
    \centering
    \includegraphics[width=\textwidth]{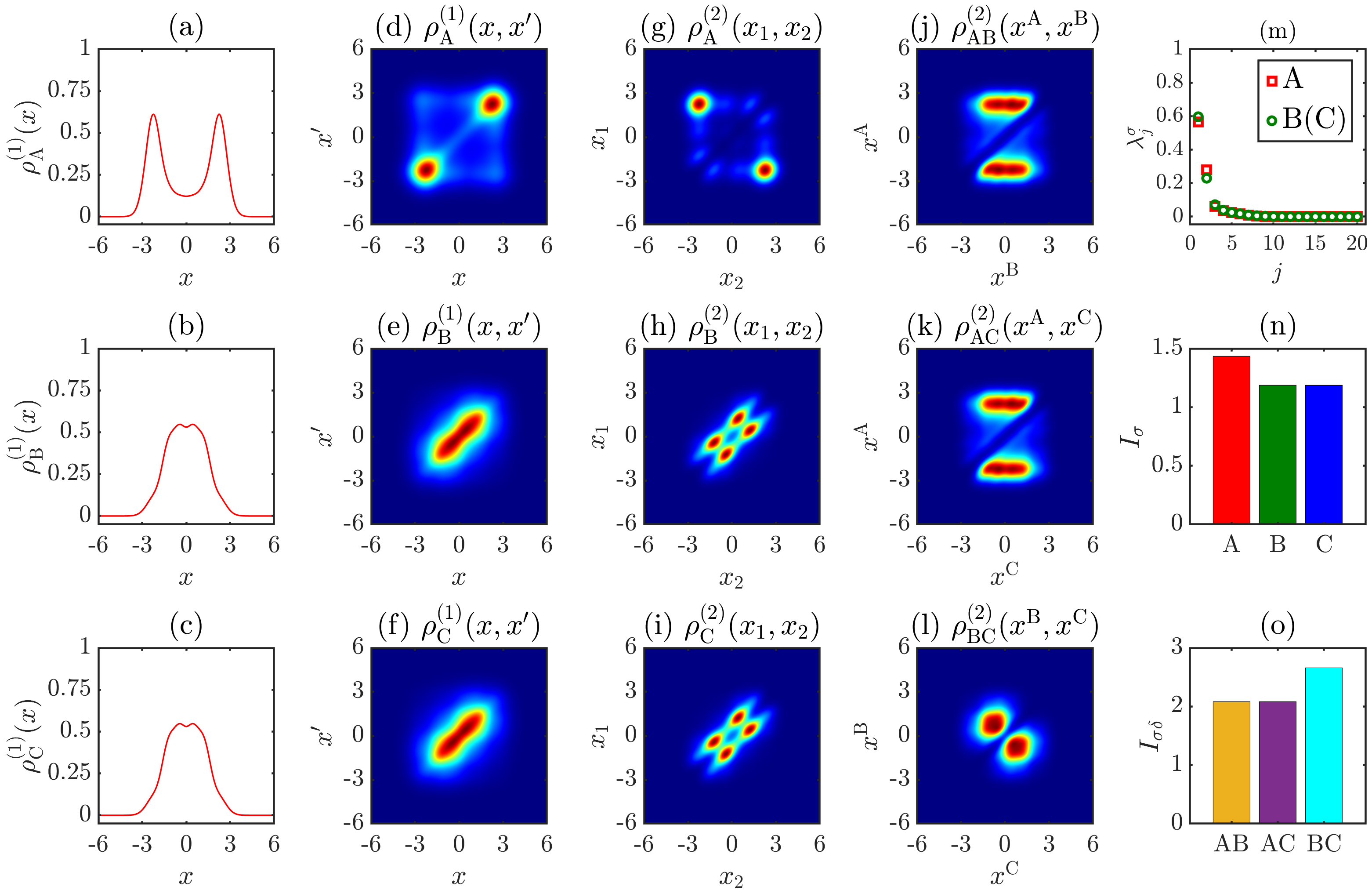}
    \caption{The quantities of interest of the many-boson ground state between vertices \textbf{1} and \textbf{4} at $g=10$ in the crossover  (specifically $g_{\rm{A}}=g_{\rm{AB}}=g_{\rm{AC}}=g_{\rm{BC}}=20,g_{\rm{C}}=g_{\rm{B}}=10 $).  (a-c) The one-body density distribution function $\rho_\sigma^{(1)}(x)$. (d-f) The one-body density matrix $\rho_\sigma^{(1)}(x,x^\prime)$. (g-i) The intraspecies two-body correlation function $\rho^{(2)}_{\sigma}(x_1,x_2)$. (j-l) The interspecies two-body correlation function $\rho^{(2)}_{\sigma\delta}(x^{\sigma},x^{\delta})$. (m) The eigenvalues $\lambda^\sigma_j$ of the one-body density matrices. (n) The intraspecies mutual information $I_\sigma$. (o) The interspecies mutual information $I_{\sigma\delta}$. \TT{In panels (d)-(l), the color gradient ranges from minimum (blue) to maximum (dark red).}} 
    \label{fig:strange_state}
\end{figure*}

We also note that around the crossover region, at $g\approx10$, where all species have a large overlap with one another \TTT{(see Fig.~\ref{fig:phase_transition_1}(f))}, all pairs of particles have approximately the same mutual information (see Fig.~\ref{fig:phase_transition_1}(d) and (e)), indicating pair correlations are spread equally among all components. This saturation of two-body correlations means that three-body correlations between all the components effectively vanish throughout the crossover. \TTT{To understand the system at this crossover point in more detail we show in Fig.~\ref{fig:strange_state} the correlation functions associated with the system at $g=10$, where the symmetry between species B and C is restored. The strong intraspecies repulsion between the A particles ensure they are mostly separated to the right and left of species B and C as shown in the one-body density distributions depicted in Figs.~\ref{fig:strange_state}(a-c), however they are not completely phase separated as there is a significant overlap between all species at the trap center. The A particles therefore possess a degree of off-diagonal long range order as can be seen in the one-body density matrix $\rho^{(1)}_\text{A}(x,x^\prime)$ in Fig.~\ref{fig:strange_state}(d). This is also echoed in its two-body correlation function $\rho^{(2)}_\text{A}(x_1,x_2)$ in Fig.~\ref{fig:strange_state}(g), which shows the A-type bosons are mostly separated from one another and hence are anti-bunched, however, some bunching correlations are also present as the particles are partially delocalized over the length of the trap. In comparison, the comparatively weakly interacting B and C species ($g_\text{B}=g_\text{C}=10$) are localized in the trap center and exhibit intraspecies bunching (see Figs.~\ref{fig:strange_state}(h) and (i)), while the stronger interspecies interactions $g_\text{BC}=20$ ensure that B and C particles have anti-bunching correlations with respect to one another (see Fig.~\ref{fig:strange_state}(l)). Importantly, we note that the arrangement of the particles and their correlations are unique, and not directly equivalent to any of the phases presented in Fig.~\ref{fig:ice-cube}. This is due to the large overlap between all the species and the presence of strong inter- and intraspecies correlations, which hints at the rich amount of non-trivial states that can be found between vertices of the phase cubes.

Finally, we remark on the the energy spectrum of the low-lying excited states for the presented model in Fig.~\ref{fig:phase_transition_1}(k). As can be clearly seen, the energy spectrum exhibits a number of avoided crossings between the low-lying even-parity eigenstates, in particular between the ground state and the even-parity first excited state. It is crucial to note that although the energy gap between the even-parity first and second excited states is relatively small at the point $g=10$, on the order of $10^{-3}$, they still do not cross.  Overall, we can infer that the complexity of the energy spectrum with the presence of these close avoided crossings and the strong correlations when interactions are finite suggest that driving the system adiabatically could be a major challenge.}

\section{Conclusion and Outlook}
\label{sec:conclusion}
In this work we have systematically laid the foundations for understanding quantum correlations, coherence and spatial localization of one-dimensional three-species mixtures of ultra-cold few bosons confined harmonically. Our calculations are based on the improved Exact Diagonalization scheme that efficiently solves the many-body Schr\"{o}dinger equation of mixtures of a few interacting indistinguishable particles in a truncated Hilbert space. This numerical tool has allowed us to calculate the full many-body ground-state wavefunction and thus explore all possible quantum correlations, coherence, spatial self-localization and entanglement. From this insight we have categorized all phases by their inter- and intraspecies coupling strengths, focusing on the limits of either the ideal ($g=0$) or the hard-core ($g\to\infty$) behavior. We have found ten ground-state phases that are unique to three-species mixtures of interacting bosons, three of which are discussed in the main text and the remaining seven in the Appendix for completeness. \TT{We have shown that we can group these phases according to their exchange symmetry, allowing us to describe analogous states to those found in two component systems, such as when we consider isotropic interspecies interactions and find the Triple Composite Fermionization and Full Fermionization phases. It is worth mentioning that the Full Fermionization phase (see Appendix~\ref{TFF}) has SU(3) symmetry and therefore can be mapped onto an effective spin model as it has been done in binary mixtures~\cite{musolino2024symmetry,barfknecht2018effects,yang2016effective,massignan2015magnetism,aupetit2022exact}.}

\TT{By removing one interspecies interactions we find a collection of exotic phases which have no straightforward analog in two-component systems. Here we have focused on the cases where there exists no invariant exchange symmetry, yielding complex states where the self-organized localization of the particles is strongly affected by correlation effects. Furthermore, we have also discussed the crossover between two of these states by simultaneously changing the intraspecies interactions.} The results show that although the correlations can be exchanged between two species, it strongly depends on the particle-particle interactions. Some correlation exchanges between two species may be more difficult to obtain in practice since the system needs driving through a quantum-matter barrier formed by the third component. This naturally gives rise to a fundamental and interesting question about the design of geodesic paths for driving quantum many-body systems to the desired state that will be addressed in future work \cite{sivak2012thermodynamic,tomka2016geodesic,kahan2019driving}. In addition to the above results it would be interesting to study mass- or particle-imbalanced systems, transitions between different correlated phases and also non-equilibrium dynamics along with the investigation of the emergence of quantum chaos. Finally, it will be interesting to derive a variational ansatz for each of the exotic phases that are unique to the three-species strongly-correlated system, \TT{which will also be useful for Monte Carlo simulations with larger particle numbers.}

\begin{acknowledgments}
The work is supported by the Okinawa Institute of Science and Technology Graduate University (OIST). All numerical simulations were performed on the high-performance computing cluster provided by the Scientific Computing and Data Analysis section at OIST. T.B., T.F., and T.D.A.-T. acknowledge support from JST COI-NEXT Grant No. JPMJPF2221 and T.F. is also supported by JSPS KAKENHI Grant No. JP23K03290. T.D.A.-T. thanks the Pure and Applied Mathematics University Research Institute at the Polytechnic University of Valencia for their kind hospitality and also acknowledges support through a Dodge Postdoctoral Researcher Fellowship at the University of Oklahoma. M.A.G.-M is supported by the Ministry for Digital Transformation and of Civil Service of the Spanish Government through the QUANTUM ENIA project call - Quantum Spain project, and by the European Union through the Recovery, Transformation and Resilience Plan - NextGenerationEU within the framework of the Digital Spain 2026 Agenda: also from Projects of MCIN with funding from European Union NextGenerationEU (PRTR-C17.I1) and by Generalitat Valenciana, with Ref. 20220883 (PerovsQuTe) and COMCUANTICA/007 (QuanTwin), and Red Tematica RED2022-134391-T. 
\end{acknowledgments}

\appendix

\section{Additional Tri-correlated States}
\label{appendix}
 
\subsection{Triple Composite Fermionization}
\label{TCF}
\begin{figure*}[tb]
    \centering
    \includegraphics[width=\textwidth]{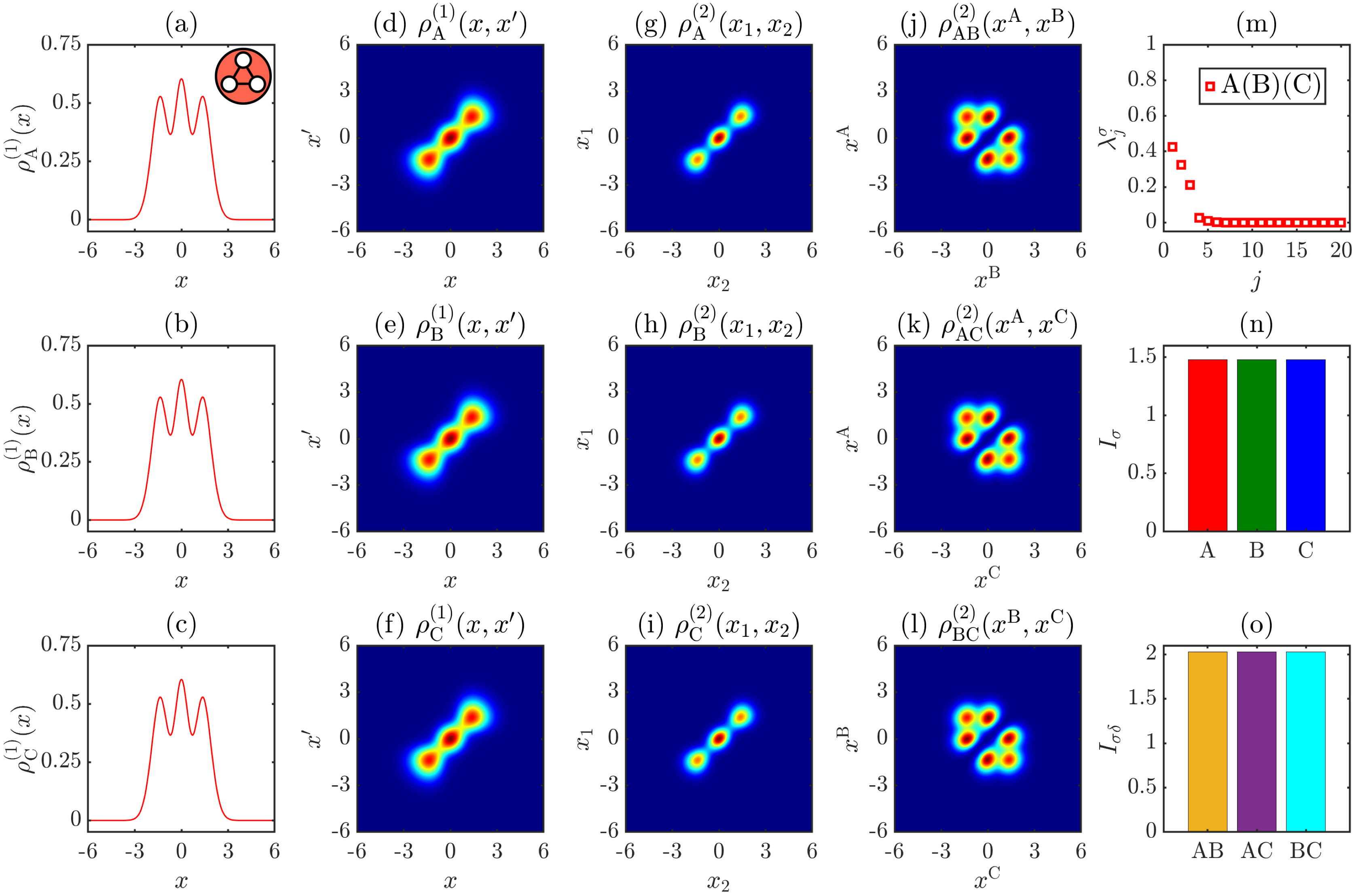}
    \caption{The ``Triple Composite Fermionization" phase ($g_{\rm{AB}}=g_{\rm{BC}}=g_{\rm{AC}}\to\infty$, and $g_{\rm{A}}=g_{\rm{B}}=g_{\rm{C}}=0$). (a-c) The one-body density distribution function $\rho_\sigma^{(1)}(x)$. (d-f) The one-body density matrix $\rho_\sigma^{(1)}(x,x^\prime)$. (g-i) The intraspecies two-body correlation function $\rho^{(2)}_{\sigma}(x_1,x_2)$. (j-l) The interspecies two-body correlation function $\rho^{(2)}_{\sigma\delta}(x^{\sigma},x^{\delta})$. (m) The eigenvalues $\lambda^\sigma_j$ of the one-body density matrices. (n) The intraspecies mutual information $I_\sigma$. (o) The interspecies mutual information $I_{\sigma\delta}$. \TT{In panels (d)-(l), the color gradient ranges from minimum (blue) to maximum (dark red).}} 
    \label{fig:triple_composite_fermionization}
\end{figure*}

\TT{When all three interspecies coupling strengths $(g_\text{AB},g_\text{AC},g_\text{BC})$ tend to infinity while all intraspecies $(g_\text{A},g_\text{B},g_\text{C})$ vanish, the system is in the ``Triple Composite Fermionization'' phase.} Since all species strongly repel each other, this phase resembles some features of composite fermionization introduced for two-component bosonic mixtures~\cite{sascha2008composite,hao2009ground1,garcia2013sharp}. The detailed results for this case are depicted in Fig.~\ref{fig:triple_composite_fermionization}, and one can immediately see that the quantities of interest are identical for all species due to the symmetries in the coupling strengths, \TT{with the groundstate being triply degenerate}. The strongly repulsive interspecies interactions result in a one-body spatial density profiles that has three peaks for each component, with the highest probability of finding a $\sigma$-species boson at the center of the trap. By looking at the reduced one-body density matrix (Figs.~\ref{fig:triple_composite_fermionization}(d-f)) one can see that the probability of a $\sigma$-species boson at the position $x$ immediately being measured at the position $x^\prime$ mainly distributes along the diagonal $x=x^\prime$, which indicates the absence of off-diagonal long-range order, which is typical for a fermionized state. Furthermore, the two body correlations functions (see panels (g-i) in Fig.~\ref{fig:triple_composite_fermionization}) show that two atoms of different species cannot be found at same position since $\rho^{(2)}_{\sigma\delta}(x^\sigma,x^\delta)$ is zero along the diagonal owing to the infinitely repulsive interspecies interaction. However two atoms of the same species can be found at the same place, either at the center of trap or slightly displaced to the left or right (Figs.~\ref{fig:triple_composite_fermionization}(j-l)). For all three components the OBDM has a number of finite eigenvalues, indicating that neither is totally condensed. Finally, since all interspecies coupling strengths are in the strongly repulsive regime, the species have the same inter- and intraspecies correlations (see Figs.~\ref{fig:triple_composite_fermionization}(n,o)). 
 
\subsection{Full Fermionization}
\label{TFF}
\begin{figure*}[tb]
    \centering
    \includegraphics[width=\textwidth]{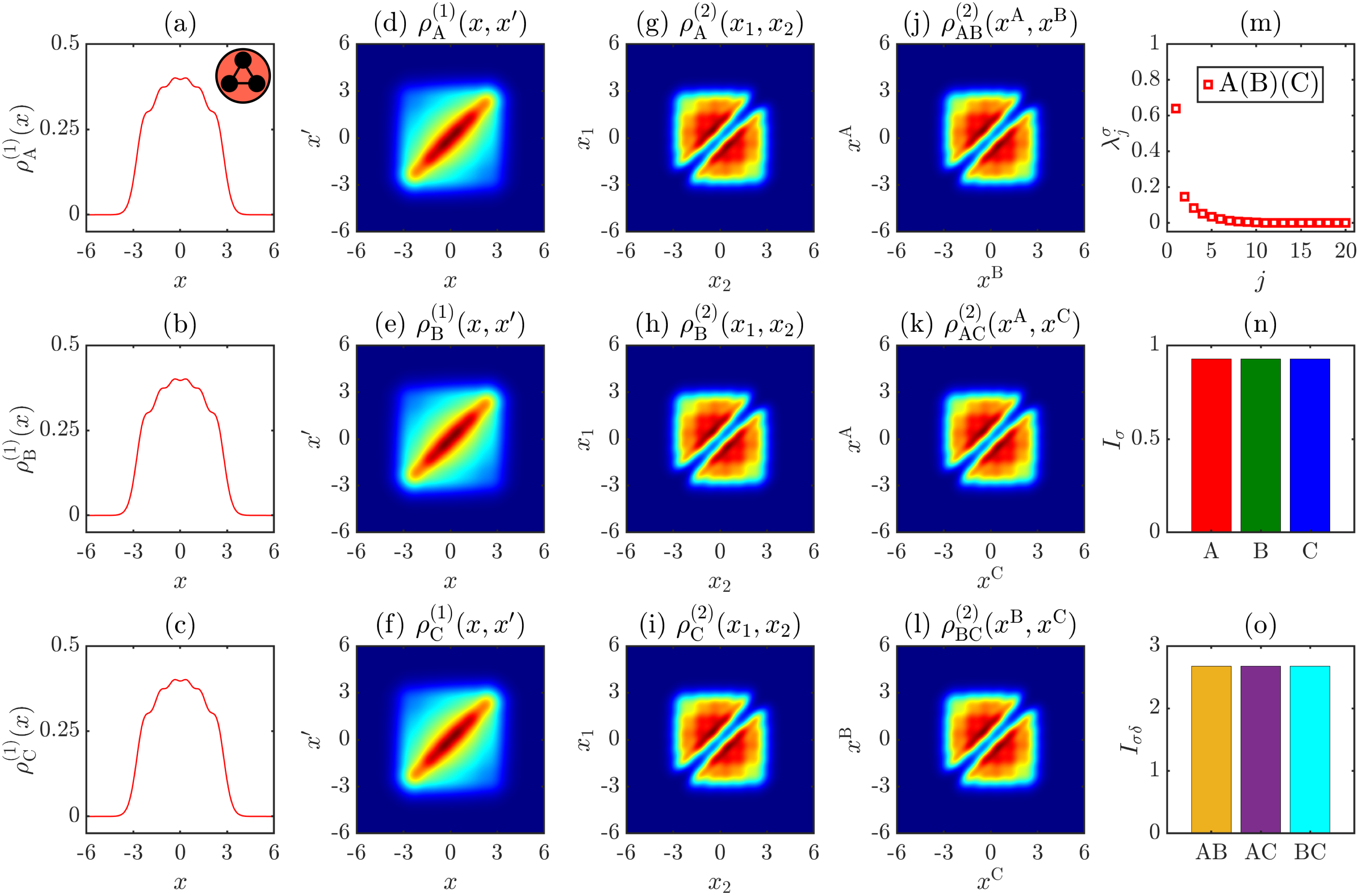}
    \caption{The ``Full Fermionization" phase ($g_{\rm{A}}=g_{\rm{B}}=g_{\rm{C}}=g_{\rm{AB}}=g_{\rm{BC}}=g_{\rm{AC}}\to\infty$). (a-c) The one-body density distribution function $\rho_\sigma^{(1)}(x)$. (d-f) The one-body density matrix $\rho_\sigma^{(1)}(x,x^\prime)$. (g-i) The intraspecies two-body correlation function $\rho^{(2)}_{\sigma}(x_1,x_2)$. (j-l) The interspecies two-body correlation function $\rho^{(2)}_{\sigma\delta}(x^{\sigma},x^{\delta})$. (m) The eigenvalues $\lambda^\sigma_j$ of the one-body density matrices. (n) The intraspecies mutual information $I_\sigma$. (o) The interspecies mutual information $I_{\sigma\delta}$. \TT{In panels (d)-(l), the color gradient ranges from minimum (blue) to maximum (dark red).}}
    \label{fig:triple_full_fermionization}
\end{figure*}

\TT{The case in which all interactions are strongly repulsive and are totally symmetric in all species is termed ``Full Fermionization" and in Fig.~\ref{fig:triple_full_fermionization} we show the numerical results for this case.} Due to the full symmetry between all six bosons, they can be found in any order starting from left to right, which results in six peaks in the density profiles depicted in Figs.~\ref{fig:triple_full_fermionization}(a-c). It can be understood as a TG gas of six infinitely repulsive bosons which can not be located at the same position as visible in the TBCFs (see Figs.~\ref{fig:triple_full_fermionization}(i-o)). In fact the results in this case straightforwardly extend the Full Fermionization case in bosonic binary mixture presented in Refs.~\cite{sascha2008composite,garcia2014quantum}. Similar to the ``Triple Composite Fermionization'' case, each species and each combination possess the same amount of intra- and interspecies correlations as $I_\text{A}=I_\text{B}=I_\text{C}$ and $I_\text{AB}=I_\text{AC}=I_\text{BC}$. 

\subsection{Induced Composite Fermionization - Phase Separation}
\label{DICFPS}
\begin{figure*}[tb]
    \centering
    \includegraphics[width=\textwidth]{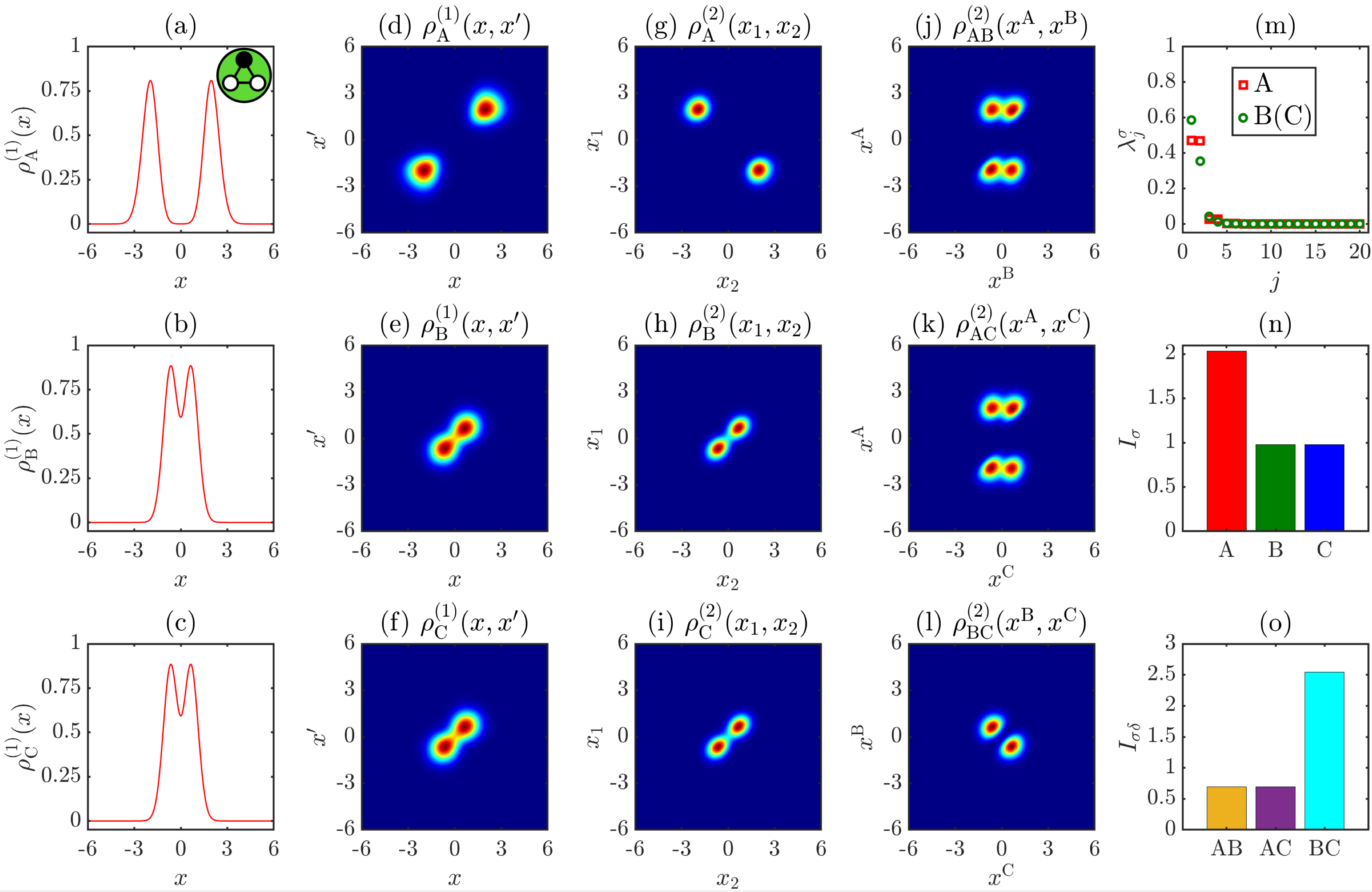}
    \caption{The ``Induced Composite Fermionization - Phase Separation" phase ($g_{\rm{A}}=g_{\rm{AB}}=g_{\rm{BC}}=g_{\rm{AC}}\to\infty$, and $g_{\rm{B}}=g_{\rm{C}}=0$). (a-c) The one-body density distribution function $\rho_\sigma^{(1)}(x)$. (d-f) The one-body density matrix $\rho_\sigma^{(1)}(x,x^\prime)$. (g-i) The intraspecies two-body correlation function $\rho^{(2)}_{\sigma}(x_1,x_2)$. (j-l) The interspecies two-body correlation function $\rho^{(2)}_{\sigma\delta}(x^{\sigma},x^{\delta})$. (m) The eigenvalues $\lambda^\sigma_j$ of the one-body density matrices. (n) The intraspecies mutual information $I_\sigma$. (o) The interspecies mutual information $I_{\sigma\delta}$. \TT{In panels (d)-(l), the color gradient ranges from minimum (blue) to maximum (dark red).}}
    \label{fig:double_induced_composite_fermionization_phase_separation}
\end{figure*}

An intriguing case occurs when one of the intraspecies interactions and all three interspecies interactions are large. \TT{We refer to this as the ``Induced Composite Fermionization - Phase Separation'' phase and it can appear in three different ways. We also remark that the system is invariant when two species without intraspecies interactions exchange. The quantities of interest for the case, in which $g_{\rm{A}}=g_{\rm{AB}}=g_{\rm{BC}}=g_{\rm{AC}}\to\infty$, $g_{\rm{B}}=g_{\rm{C}}=0$, are shown in Fig.~\ref{fig:double_induced_composite_fermionization_phase_separation}.} One can immediately see from Fig.~\ref{fig:double_induced_composite_fermionization_phase_separation}(a) that the infinitely repulsive intraspecies interactions between the A bosons leads to them being separated into two equal parts and localized at the edges. Furthermore, the lack of off-diagonal terms in all OBDMs visible in Figs.~\ref{fig:double_induced_composite_fermionization_phase_separation}(d-f), shows the absence of the long-range correlations within each species. These OBDMs show that one can only find bosons of kind A either on the left or right of mostly centered B and C bosons, and consequently $\rho_\text{A}^{(1)}(x,x^\prime)$ is fragmented with two doubly-degenerate eigenvalues close to $0.5N_\text{A}$ as depicted in Fig.~\ref{fig:double_induced_composite_fermionization_phase_separation}(m). 

The two-body correlations between the two A bosons, Fig.~\ref{fig:double_induced_composite_fermionization_phase_separation}(g), show that the particles are anti-bunched, meaning that one will always find one A atom on one side of the trap, and the other A atom on the other side. Meanwhile the atoms of the other two species are localized at the center of the trap and exhibit some features similar to the Composite Fermionization phase in binary mixtures~\cite{sascha2008composite}. In particular two identical (B or C species) bosons can sit on top each other, whereas a C boson and a B boson avoid occupying the same place. This is due to the correlations with the A species and the eigenvalues of the OBDM for components B and C \TT{show the occupation of higher natural orbitals}, confirming the absence of coherence. The two-body correlation functions for species B and C, Figs.~\ref{fig:double_induced_composite_fermionization_phase_separation}(h,i), show that the two bosons of each of these species have a high probability to occupy the same space due to their vanishing intraspecies interactions. However, due to the repulsion between the components, each one is located slightly away from the center of the trap, which can also be clearly seen in the two-body correlations between an A boson and a B (or C) boson in Figs.~\ref{fig:double_induced_composite_fermionization_phase_separation}(j,k). Since the interspecies interaction between B and C is large and repulsively, they phase separate with the two B atoms being either on the left or the right of the two  C atoms, see Fig.~\ref{fig:double_induced_composite_fermionization_phase_separation}(l). Finally, since species A is split, it has a larger amount of intraspecies correlations than species B(C), $I_\text{A}>I_\text{B}=I_\text{C}$, but the components B and C exhibit the highest level of interspecies correlations. 

\subsection{Phase Separation}
\label{PS} 
\begin{figure*}[tb]
    \centering
    \includegraphics[width=\textwidth]{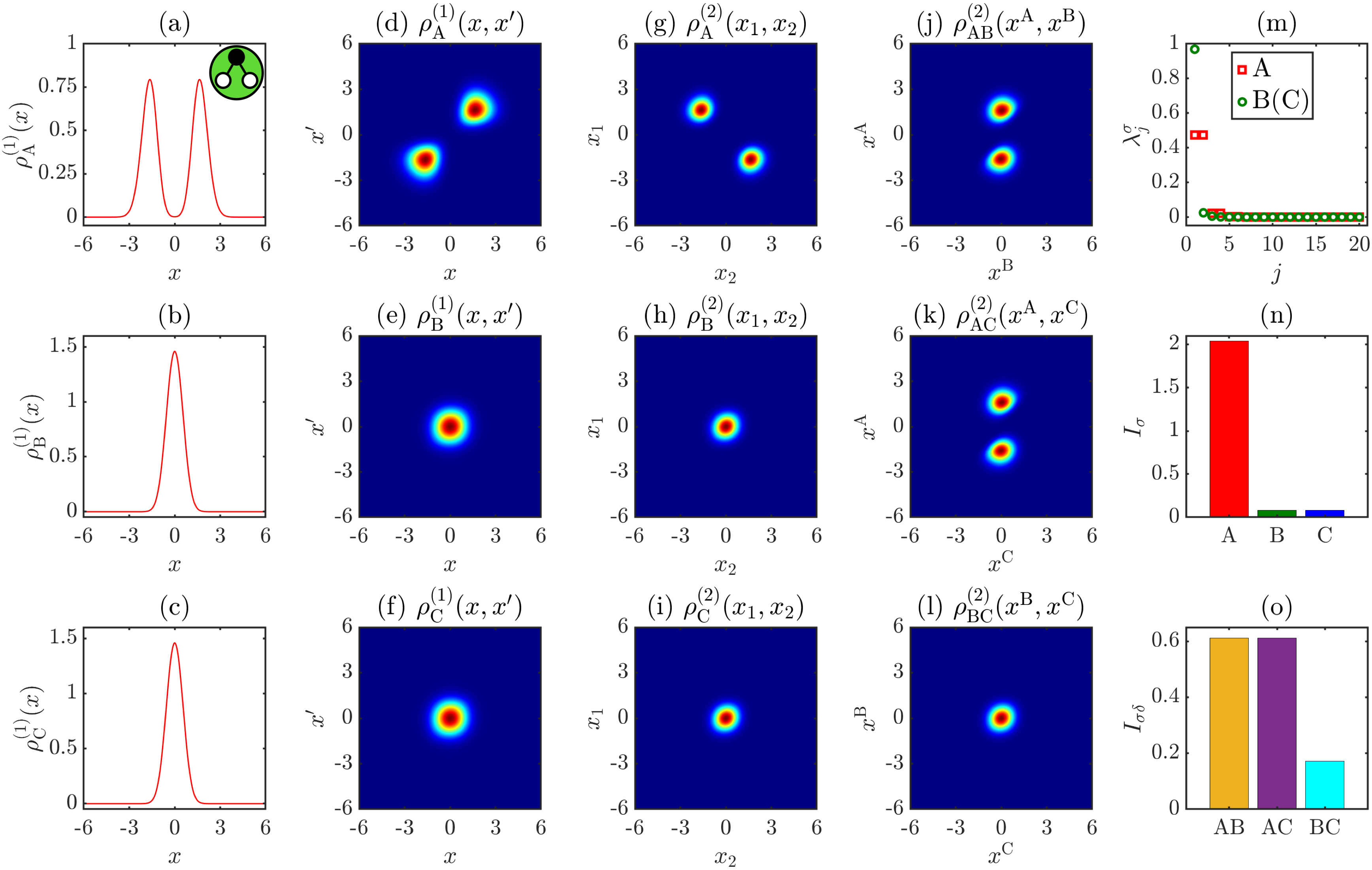}
    \caption{The ``Phase Separation" phase ($g_{\rm{A}}=g_{\rm{AB}}=g_{\rm{AC}}\to\infty$ and $g_{\rm{B}}=g_{\rm{C}}=g_{\rm{BC}}=0$). (a-c) The one-body density distribution function $\rho_\sigma^{(1)}(x)$. (d-f) The one-body density matrix $\rho_\sigma^{(1)}(x,x^\prime)$. (g-i) The intraspecies two-body correlation function $\rho^{(2)}_{\sigma}(x_1,x_2)$. (j-l) The interspecies two-body correlation function $\rho^{(2)}_{\sigma\delta}(x^{\sigma},x^{\delta})$. (m) The eigenvalues $\lambda^\sigma_j$ of the one-body density matrices. (n) The intraspecies mutual information $I_\sigma$. (o) The interspecies mutual information $I_{\sigma\delta}$. \TT{In panels (d)-(l), the color gradient ranges from minimum (blue) to maximum (dark red).}}
    \label{fig:double_phase_separation}
\end{figure*}

\TT{The situation of one component with a strong intraspecies coupling and strong interspecies coupling with the other two components defines the ``Phase Separation'' case and in the following we present the results with interaction strengths $g_{\rm{A}}=g_{\rm{AB}}=g_{\rm{AC}}\to\infty$ and $g_{\rm{B}}=g_{\rm{C}}=g_{\rm{BC}}=0$. }Due to the symmetry between the B and the C component, the distribution of the particles obeys the same logic as in the Phase Separation case for Bose-Bose mixture~\cite{garcia2014quantum}. From the plots in Figs.~\ref{fig:double_phase_separation}(a-c), one can see that species B and C, whose intraspecies interactions vanish, are confined to the center of the trap and remain coherent with the occupancy of their highest natural orbitals roughly being $0.97$. In contrast, species A is split into two equal parts that are located towards the edges, with zero probability of being found in the center. This results in doubly-degenerated natural orbital occupation numbers of the A species, with both of them close to $0.5$. Splitting of the A species due to its strong intraspecies repulsion can be seen in Fig.~\ref{fig:double_phase_separation}(g), whilst the B and C bosons are located on top each other at the trap center depicted in Figs.~\ref{fig:double_phase_separation}(h,i,l). These findings are further emphasized in the AB and AC two-body correlation functions, where the likelihood of finding A-type bosons around $x^\text{A}=0$ is fully absent, but two large peaks aligned along $x^{\text{B}(\text{C})}=0$  can be seen in Figs.~\ref{fig:double_phase_separation}(j,k). Although they do not directly interact with each other, species B and C can be seen to retain a small level of intraspecies correlations induced via the couplings to species A as $I_\text{B}=I_\text{C}\neq 0$. The infinitely repulsive intraspecies and interspecies interactions with the B and C species leads to the large value of $I_\text{A}$, indicating that species A has very strong intraspecies correlations. In terms of interspecies correlations, the components A and B, and A and C \TT{are comparable}, which is larger than the one of the components B and C as seen in Fig.~\ref{fig:double_phase_separation}(o).

\subsection{Correlation-induced Bunching type II} 
\label{CIBI}
\begin{figure*}[t!b]
    \centering
    \includegraphics[width=\textwidth]{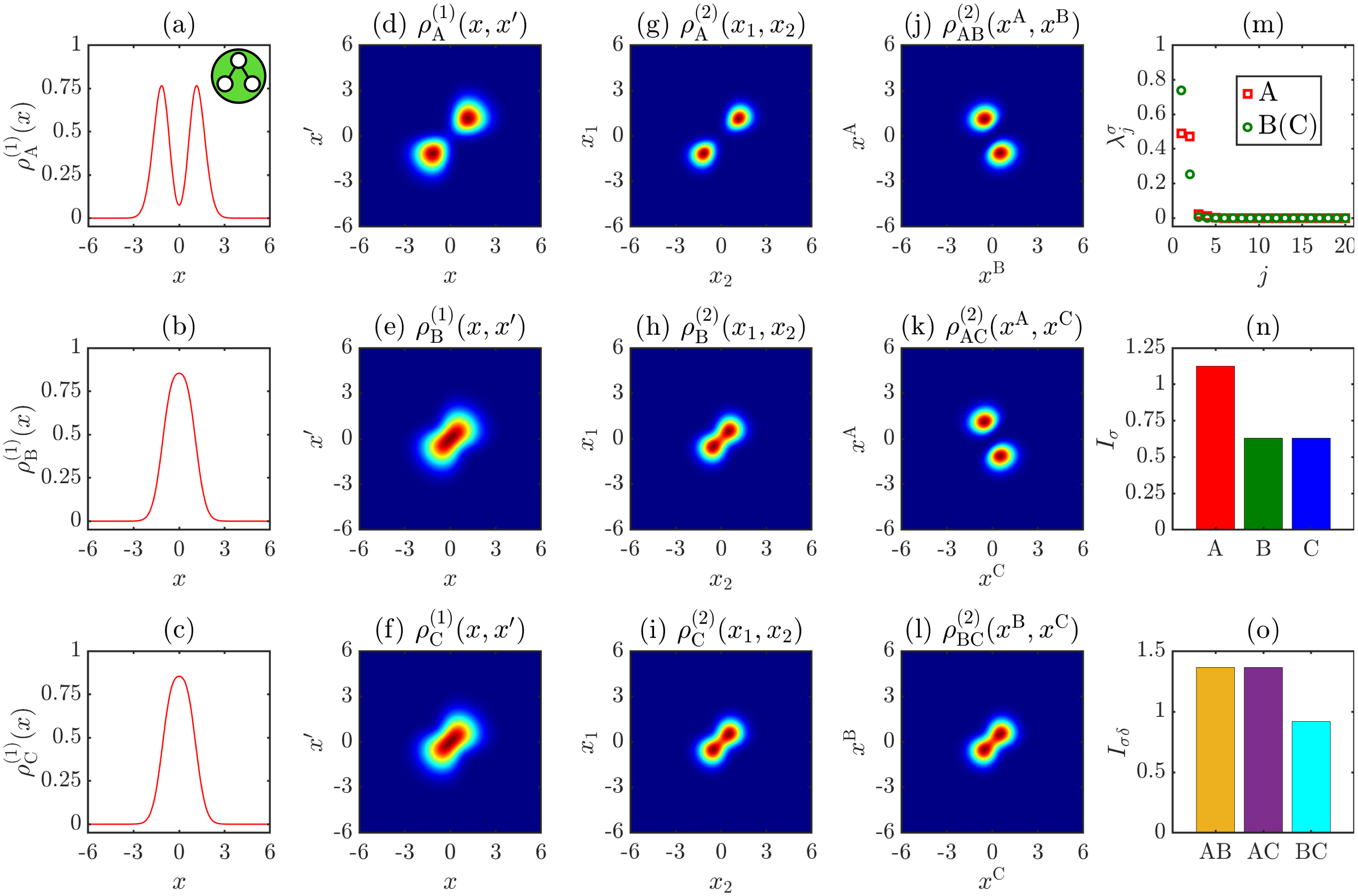}
    \caption{The ``Correlation-induced Bunching type II" phase ($g_{\rm{AB}}=g_{\rm{AC}}\to\infty$, and $g_{\rm{A}}=g_{\rm{B}}=g_{\rm{C}}=g_{\rm{BC}}=0$). (a-c) The one-body density distribution function $\rho_\sigma^{(1)}(x)$. (d-f) The one-body density matrix $\rho_\sigma^{(1)}(x,x^\prime)$. (g-i) The intraspecies two-body correlation function $\rho^{(2)}_{\sigma}(x_1,x_2)$. (j-l) The interspecies two-body correlation function $\rho^{(2)}_{\sigma\delta}(x^{\sigma},x^{\delta})$. (m) The eigenvalues $\lambda^\sigma_j$ of the one-body density matrices. (n) The intraspecies mutual information $I_\sigma$. (o) The interspecies mutual information $I_{\sigma\delta}$. \TT{In panels (d)-(l), the color gradient ranges from minimum (blue) to maximum (dark red).}}
    \label{fig:cib_phase_separation_type_I}
\end{figure*}

The case in which only two infinitely repulsive interspecies interactions exist is termed ``Correlation-induced Bunching type II''.  In Fig.~\ref{fig:cib_phase_separation_type_I} we show the realization corresponding to the interaction configuration $g_{\rm{AB}}=g_{\rm{AC}}\to\infty$, and $g_{\rm{A}}=g_{\rm{B}}=g_{\rm{C}}=g_{\rm{BC}}=0$. This case is dominated by two insights. First, any overlap of component A with either of the other two is energetically costly, however components B and C can overlap. One can see from Figs.~\ref{fig:cib_phase_separation_type_I}(a,d) that this leads to component A being split into two parts with its one-body density matrix having two dominant eigenvalues close to $0.5$. Additionally, one can see from Fig.~\ref{fig:cib_phase_separation_type_I}(g) that $\rho^{(2)}_\text{A}(x_1,x_2)$ is concentrated along the diagonal, which means that the two A bosons are bunched at one side of the trap. This is due to the presence of a mediated attractive interaction through the B and C components which acts to bind the A particles together in the absence of intraspecices coupling $g_\text{A}=0$. Second, even though components B and C do not directly interact, they indirectly interact via their respective interactions with component A. This can be confirmed from the fact that their OBDMs possess two large eigenvalues, despite both components showing a Gaussian-like density profile localized about the center of the trap. This is consistent with the narrowing of the one-body correlation functions $\rho^{(1)}_\text{B}(x,x^\prime)$ and $\rho^{(1)}_\text{C}(x,x^\prime)$ along the off-diagonal $x=-x^\prime$. These mediated interactions are also attractive as seen in the particle bunching along the diagonal of the two-body correlation functions $\rho^{(2)}_\text{B}(x_1,x_2)$, $\rho^{(2)}_\text{C}(x_1,x_2)$ and $\rho^{(2)}_\text{BC}(x^\text{B},x^\text{C})$, which are identical due to symmetry between the components. Similar correlation effects  are manifested in the two-body correlation functions $\rho^{(2)}_\text{AB}(x^\text{A},x^\text{B})$ and $\rho^{(2)}_\text{AC}(x^\text{A},x^\text{C})$ by the peaks being slightly off from  $x^{\rm{B}(\rm{C})} = 0$, and all induced correlations can be quantified by the non-zero values of $I_\text{A}$, $I_\text{B}$, $I_\text{C}$ and $I_\text{BC}$ as shown in Figs.~\ref{fig:cib_phase_separation_type_I}(n,o).

\subsection{Correlation-induced Bunching type III}
\label{CIBIII} 

\begin{figure*}[tb]
    \centering
    \includegraphics[width=\textwidth]{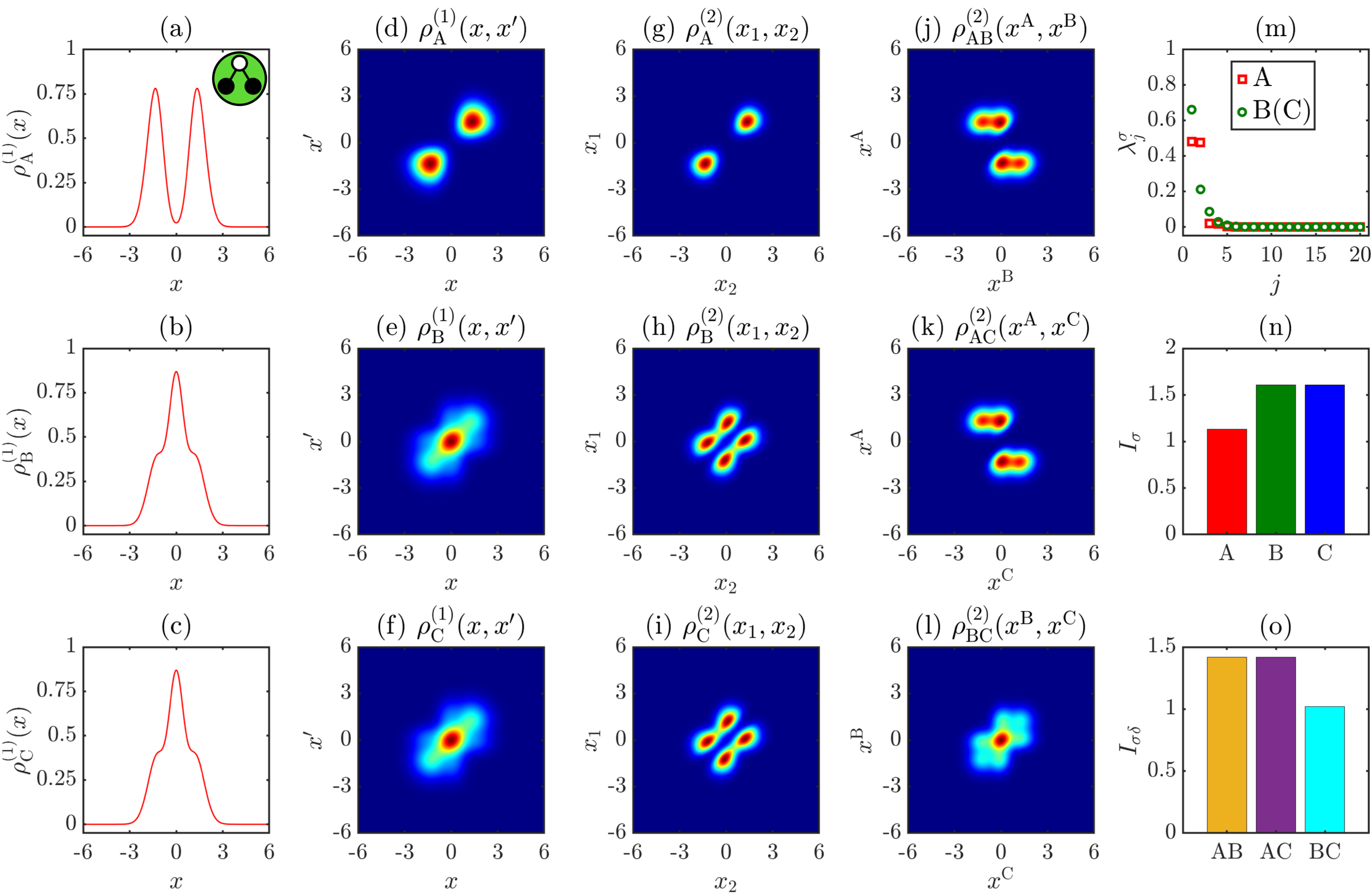}
    \caption{The ``Correlation-induced Bunching type III" phase ($g_{\rm{B}}=g_{\rm{C}}=g_{\rm{AB}}=g_{\rm{AC}}\to\infty$ and $g_{\rm{A}}=g_{\rm{BC}}=0$). (a-c) The one-body density distribution function $\rho_\sigma^{(1)}(x)$. (d-f) The one-body density matrix $\rho_\sigma^{(1)}(x,x^\prime)$. (g-i) The intraspecies two-body correlation function $\rho^{(2)}_{\sigma}(x_1,x_2)$. (j-l) The interspecies two-body correlation function $\rho^{(2)}_{\sigma\delta}(x^{\sigma},x^{\delta})$. (m) The eigenvalues $\lambda^\sigma_j$ of the one-body density matrices. (n) The intraspecies mutual information $I_\sigma$. (o) The interspecies mutual information $I_{\sigma\delta}$. \TT{In panels (d)-(l), the color gradient ranges from minimum (blue) to maximum (dark red).}}
    \label{fig:cib_phase_separation_type_III}
\end{figure*}

\TT{In this phase two of the intraspecies interactions tend to infinity, and both of these species interact strongly with the remaining one, as illustrated in Fig.~\ref{fig:cib_phase_separation_type_III}. Here the interactions are explicitly given by $g_{\rm{B}}=g_{\rm{C}}=g_{\rm{AB}}=g_{\rm{AC}}\to\infty$ and $g_{\rm{A}}=g_{\rm{BC}}=0$.} Note that since $g_\text{B}=g_\text{C}\to\infty$ and $g_\text{BC}=0$, species B and C are symmetric and exhibit the same physics. One can therefore immediately note that species A and B(C) have similar density profiles to the Correlation-induced Bunching phase. More specifically, species B(C) occupies mostly the center of the trap but is extended towards to the edges, starting from position at which $\rho^{\text{B}(\text{C})}(x)$ is exactly equal to its half maximum. The two A-species bosons split and can be found either to the left or to the right of the central B and C cloud as seen from the intraspecies two-body correlations $\rho^{(2)}_\text{AB}(x^\text{A},x^\text{B})$ and $\rho^{(2)}_\text{AC}(x^\text{A},x^\text{C})$. Although the two B(C) atoms avoid locating at the same position as $\rho^{(2)}_\text{B(C)}(x_1,x_2=x_1) = 0$, one B-type and one C-type boson can be found at same location with high probability at the center of the parabolic trap. Overall, these three species are not totally coherent, with a few natural orbital occupancies dominantly populated. In comparison with the ``Correlation-induced Bunching" phase illustrated in Fig.~\ref{fig:cib_phase_separation_type_II}, $I_\text{C}$ is substantially enhanced indicating a higher level of intraspecies correlations as $g_\text{C}\to\infty$ and $I_\text{C}=I_\text{B}>I_\text{A}$. Furthermore, it is seen that the level of interspecies correlations in the composite AC increase and is same as the composite AB. 

\subsection{Correlation-induced Anti-bunching type II}
\label{CIABII}
\begin{figure*}[tb]
    \centering
    \includegraphics[width=\textwidth]{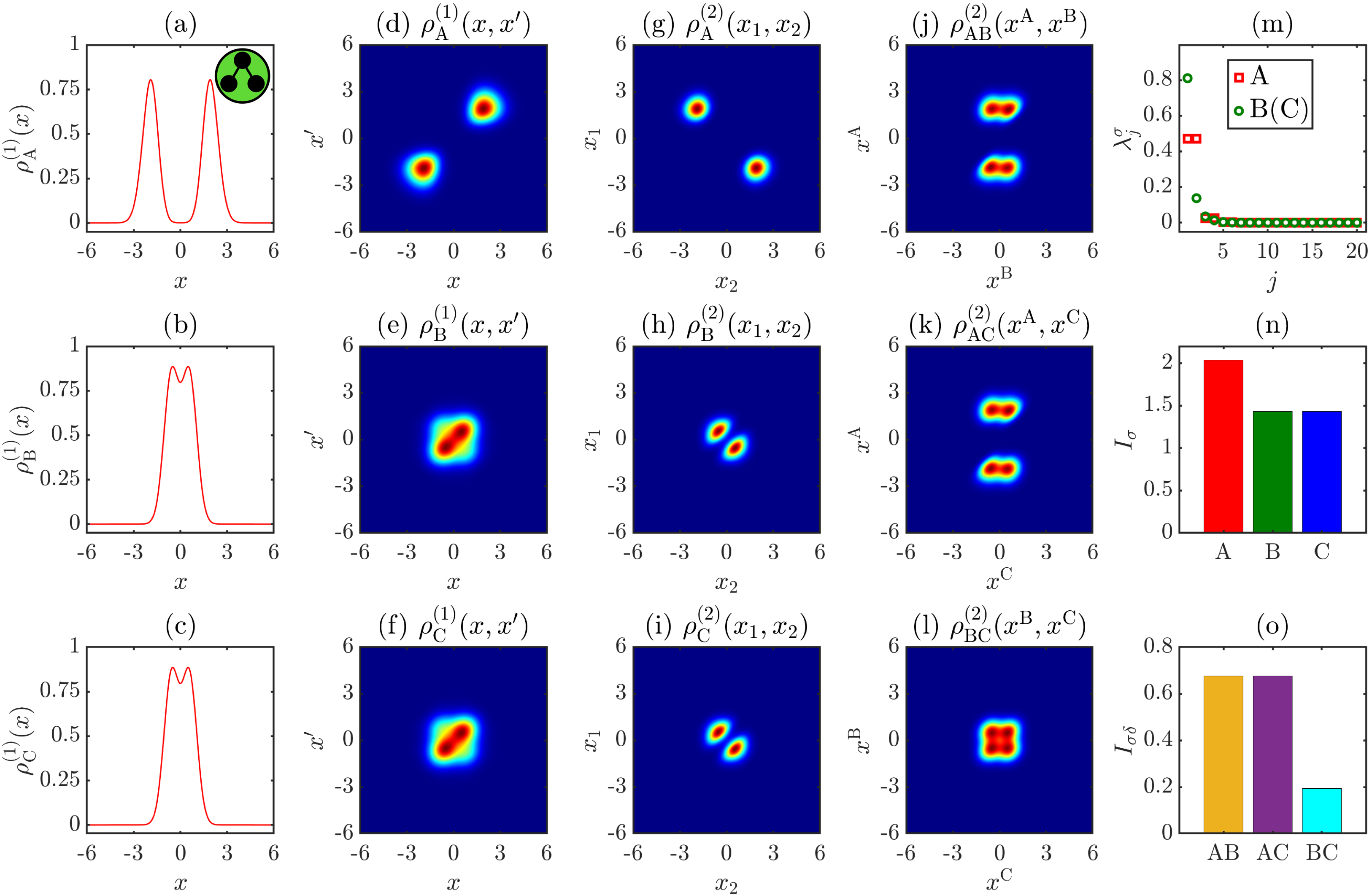}
    \caption{The ``Correlation-induced Anti-bunching type II" phase ($g_{\rm{A}}=g_{\rm{B}}=g_{\rm{C}}=g_{\rm{AB}}=g_{\rm{AC}}\to\infty$, and $g_{\rm{BC}}=0$). (a-c) The one-body density distribution function $\rho_\sigma^{(1)}(x)$. (d-f) The one-body density matrix $\rho_\sigma^{(1)}(x,x^\prime)$. (g-i) The intraspecies two-body correlation function $\rho^{(2)}_{\sigma}(x_1,x_2)$. (j-l) The interspecies two-body correlation function $\rho^{(2)}_{\sigma\delta}(x^{\sigma},x^{\delta})$. (m) The eigenvalues $\lambda^\sigma_j$ of the one-body density matrices. (n) The intraspecies mutual information $I_\sigma$. (o) The interspecies mutual information $I_{\sigma\delta}$. \TT{In panels (d)-(l), the color gradient ranges from minimum (blue) to maximum (dark red).}}
    \label{fig:double_squeezed_TG_phase_separation}
\end{figure*}

\TT{This phase is characterized by all three intraspecies interactions and two interspecies interactions being infinite and shares some similarities with the ``Correlation-induced Anti-bunching" case as can be seen in Fig.~\ref{fig:double_squeezed_TG_phase_separation}.} In the following we discuss the case, for which $g_{\rm{A}}=g_{\rm{B}}=g_{\rm{C}}=g_{\rm{AB}}=g_{\rm{AC}}\to\infty$, and $g_{\rm{BC}}=0$. One can see that the two A atoms fragment and that the central B and C clouds form two TG gases according to their OBDM and their intraspecies two-body correlation functions. The two body correlation functions $\rho^{(2)}_\text{AB}(x^\text{A},x^\text{B})$ and $\rho^{(2)}_\text{AC}(x^\text{A},x^\text{C})$ are identical and show that the two A bosons are separated, while the B and C bosons spread out due to the strong intraspecies repulsion, but still have large overlap around $x^\text{B(C)} = 0$ and can be at the same place concurrently. Furthermore, $\rho^{(2)}_\text{BC}(x^\text{B},x^\text{C})$ confirms that species B and C are spread around the center of the trap and form two independent squeezed TG gases due to the fact that there is no interspecies interactions. From Fig.~\ref{fig:double_squeezed_TG_phase_separation}(n) we can see strong intraspecies correlations in all species, particularly $I_\text{A}>I_\text{B}=I_\text{C}$. As illustrated in Fig.~\ref{fig:double_squeezed_TG_phase_separation}(o), the interspecies correlations between species A and B, as well as A and C, are equivalent and exceed those between species B and C, i.e. $I_\text{AB} = I_\text{AC} > I_\text{BC}$. 

\section{Numerical Convergence}
\label{appendixx}
\TTT{In this appendix, we detail the parameters used in the calculations and comment on the convergence of our numerical tools. In our simulations, the single-particle operators and functions are represented by the discrete variable technique (specifically, the Colbert-Miller method~\cite{colbert1992novel}) and everything is performed in a 1D box which is characterized by a spatial grid uniformly ranging from -10 to 10 with 1025 points. We employ $M_\sigma = 20$ single-particle functions and choose the energy-cutoff $E_{opt} = 50$ which results in a total of $D = 1154034$ permanents used to expand the ansatz~\eqref{eq:ansatze} after removing unnecessary permanents according to their parity symmetry. Additionally, we use the $40$ energetically lowest eigenvalues of the relative motion of two particles in a harmonic oscillator~\cite{busch1998two} to construct the two-body effective (pseudo) potential for obtaining the matrix elements $W^{\sigma}_{k\ell mn}$ and $W^{\sigma\delta}_{k\ell mn}$ as in Refs.~\cite{rotureau2013interaction,lindgren2014fermionization,dehkharghani2015quantum,anhtai2023quantum,rammelmuller2023modular,brauneis2024comparison}. We will demonstrate below that the our first-principles calculations are numerically converged within this set of parameters. 

To do this, let us demonstrate the convergence of the ground-state energy and the one-body density distribution function as $E_{opt}$ varies for the Full Fermionization phase in which all interaction strengths are equally large, $g_\sigma=g_{\sigma\delta}=20$. We use this as representative example as all the interactions are large and therefore the computational cost for calculating this many-body ground state should be the largest. It is reasonable to assume that if the results in this extreme limit are numerically exact (converged), the results for the other phases will be also converged in the same manner. Importantly, we remark that previous works~\cite{hao2009ground1,hao2009ground2,garcia2013quantum,garcia2013sharp,garcia2014quantum} have shown that a number of single-particle functions $M_\sigma = 20$ is sufficiently large for obtaining relatively well converged findings in the fully fermionized phase of Bose-Bose mixtures with the standard Exact Diagonalization scheme. Therefore, we also use $M_\sigma = 20$ single-particle functions per each species to expand the many-boson ansatz~\eqref{eq:ansatze} in this work. Nevertheless, using $M_\sigma = 20$ will result in the truncated Hilbert space with the dimension of 9,262,000 necessitating a substantial amount of memory that is not available on a single node of modern high-performance computers to the best of our knowledge. Although the parity invariance of the many-body Hamiltonian allows us to significantly reduce the dimension of the truncated Hilbert space to 4,631,000, this still requires a large amount of memory beyond what is available to us. To overcome this obstacle, as mentioned in the main text, the energy-pruning truncation technique~\cite{chrostowski2019efficient,plodzien2018numerically} is employed to remove highly-excited permanents that have an infinitesimal contribution in the many-body wavefunction for a fixed and sufficiently large number of single-particle functions. Therefore, the choice of the energy cutoff $E_{opt}$ is the sole parameter we use to control the convergence of the ab initio calculations in this work. 

We examine the convergence of our simulations with respect to $E_{opt}$ in Fig.~\ref{fig:numerical_convergence}(a), which clearly illustrates that as $E_{opt}$ increases, the ground-state energy of the system  rapidly converges. Furthermore, Fig.~\ref{fig:numerical_convergence}(b) reveals that when $E_{opt}>40$ the discrepancies between the one-body density distribution functions $\rho^{(1)}(x)$ are negligible. We can therefore infer that the quantities of interest in the ab initio calculations are numerically converged with the given set of parameters for $E_{opt}>40$. Consequently, we set $E_{opt}=50$ in all calculations to guarantee the convergence while balancing the computational resources and execution times, especially in the exploration of the crossover between phases. 

Importantly, we should mention that although the results are numerically converged, the ground-state energy and the one-body density distribution function shown in Fig.~\ref{fig:numerical_convergence} are not exactly the same as the hard-core Tonks-Girardeau limit, which can be solved by the use of the Bose-Fermi mapping theorem. This deviation is mainly due to the use of large but still finite interaction strengths ($g=20$) and that of the effective-interaction approach in our ab-initio calculations. The latter ensures that we avoid the overestimation of eigenenergies beyond the Tonks-Girardeau limit, which is influenced by sharp cusps in the many-body wavefunction resulting from the bare $\delta$-function interaction  \cite{rammelmuller2023modular,brauneis2024comparison,rotureau2013interaction,lindgren2014fermionization,dehkharghani2015quantum,rojo2022direct}, thereby yielding a better degree of convergence in the truncated Hilbert space as can be seen in Fig.~\ref{fig:numerical_convergence}. While increasing the values of $g_\sigma$ and $g_{\sigma\delta}$ further can allow us to asymptotically approach the hard-core Tonks-Girardeau limit, it would require considerable computational resources needed to obtain converged results. Nevertheless, we remark that even setting the interaction strength being very large, the numerical results would still differ from the actual hard-core TG limit as shown in Ref.~\cite{leveque2020many} (we note that the authors have used $g=1500$). Overall, we find that the choice of $g=20$ allows to qualitatively describe the strongly interacting regime and give accurate critical insights into the correlation effects and spatial localization in strongly-correlated three-species mixtures. 
\begin{figure}
    \centering
    \includegraphics[width=\linewidth]{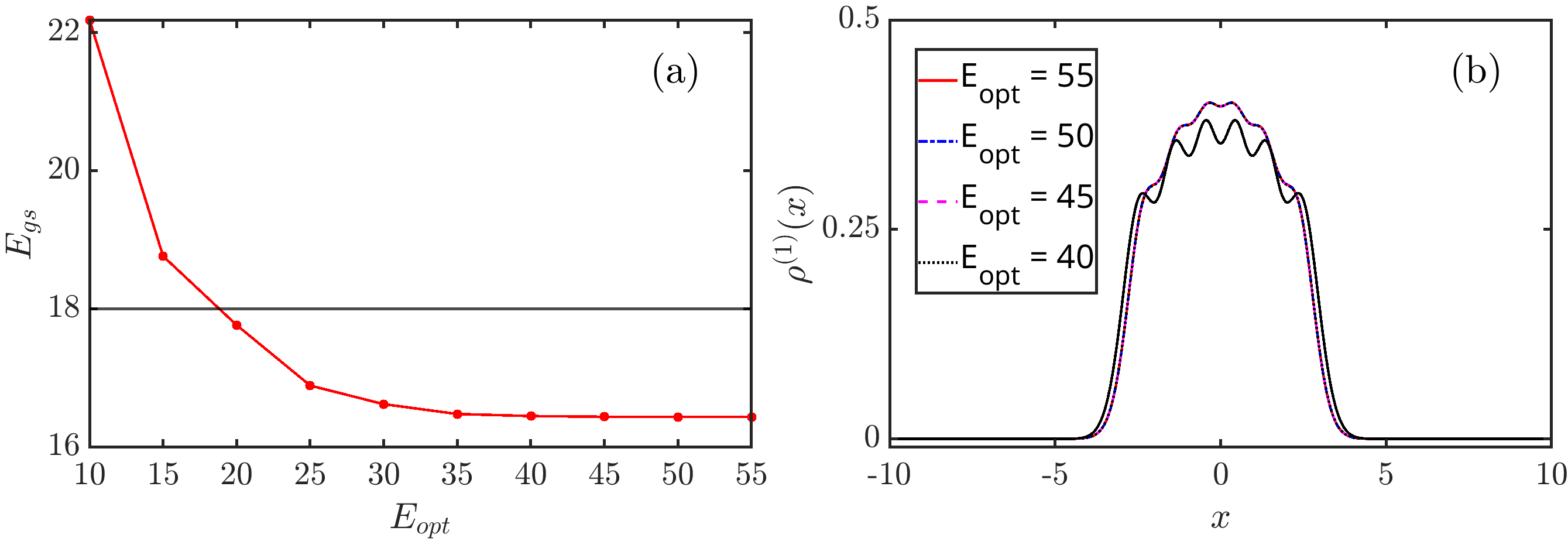}
    \caption{Convergence of the fully fermionized phase as a function of the value of the energy cut-off $E_{opt}$. Panel (a) shows the ground-state energy $E_{gs}$ and panel (b) the one-body density distribution function, $\rho^{(1)}(x)$. When all the interaction strengths are repulsively infinite, the system under consideration can be solved analytically by mapping it to six non-interacting spinless fermions confined a 1D harmonic trap via the Bose-Fermi mapping theorem~\cite{girardeau1960relationship,girardeau2001ground}. The black solid lines indicates the ground-state energy and the one-body density respectively obtained from the Bose-Fermi mapping theorem, $E_{\infty}=N^2/2=18$ and $\rho^{(1)}_{\infty}(x) = \sum\limits_{n=0}^{5}|\varphi_n^{HO}(x)|^2$ with $\varphi_n^{HO}(x)$ being the eigenfunctions of a 1D harmonic oscillator.}
    \label{fig:numerical_convergence}
\end{figure}

}
\section*{Data Availability}
All data that support the findings of this study are included within the article (and any supplementary files).

\appendix

\bibliography{apssamp}

\end{document}